# An Attempt to Improve Understanding of the Physics behind Superconductor Phase Transitions and Stability – Revised submission

Harald Reiss[#]

Department of Physics
University of Wuerzburg, Am Hubland, D-97074 Wuerzburg, FRG
harald.reiss@physik.uni-wuerzburg.de

**Abstract**
Under disturbances, superconductors may experience sudden, most undesirable phase transitions (quench) from superconducting to normal conducting state. Quench may lead to damage or even to catastrophic conductor failure. A superconductor is stable if it does not quench. By numerical simulations, the paper investigates coupling between superconductor stability and time dependence of phase transitions. For this purpose, Finite Element and Monte Carlo methods are applied to multi-filament, BSCCO 2223 and to thin film, coated YBaCuO 123 superconductors. Focus is on transient temperature distributions under multiple internal heat transfer (solid conduction and, in thin films, radiation), a suggested operator method to solve the incompleteness problem of radiative transfer, and time dependence of the order parameter obtained from a quantum-mechanical model. Explanation of the localizability of disturbances and their impacts in thin films are additional problems. These investigations shall contribute to improve understanding the physics behind the stability problem, in particular if under disturbances the material is already close to a phase transition. Traditional stability models cannot provide this information.

[#] Associate (Außerplanmäßiger) Professor

## 1    Survey: Stability of Superconductors against Quench

Under disturbances, superconductors may experience sudden, most undesirable phase transitions (quench) from superconducting to normal conducting state. Quench may lead to damage or even to catastrophic conductor failure. A superconductor is "stable" if it does not quench.

Disturbances are responsible for a variety of losses, like transformation of released mechanical to thermal energy under conductor movement induced by Lorentz forces, or absorption of high energy, particle radiation, or fault currents. Disturbances frequently are transient, but there are also permanent disturbances like flux flow losses if transport current density exceeds critical current density, or, under AC currents, hysteretic and, in multi-filamentary conductors, coupling losses. See Wilson [1], Chap. 5 for a comprehensive catalogue of superconductor disturbances.

In a rough picture, the physics behind quench is sudden conversion of stored mechanical or electromagnetic energy, the latter originating from screening and transport currents, to thermal energy in a magnetic field.

Quench can be avoided by application of stability models for design of superconductor geometry (cross section of wires or filaments, aspect ratio of thin films) and for safety during conductor manufacture, winding and operation of a superconducting device.

The literature on superconductor stability, published since about the 1980s, is very large, and a variety of standard, analytic stability models has been suggested. These models usually assume worst case conditions, or apply safety margins to make sure, from theoretical reasoning and from experience, the conductor will not quench. The models apply algebraic expressions, see Wilson [1], Dresner [2] or Seeger (Handbook of Superconductivity [3], and essentially are energy balances, compare e. g. Wilson [1], Chap. 6 to 9, and therein the derivation of stability parameters, β, Eqs. (7.7) and (7.27).

From all-day practice, application of standard stability models is successful if the superconductor is far from its critical states. But the probability that a superconductor experiences a quench increases the more, the closer the superconductor approaches its critical values of temperature, magnetic field or current density. It is *this* situation that shall be analysed in this paper.

A discussion of properties, benefits, problems and of also risks of standard stability models, in relation to numerical investigation of superconductor stability, has been presented recently, see [4-7].

Quench is a short-time physics problem; it proceeds on time scales of milliseconds and less. Numerical calculation of heat transfer and of transient temperature distribution in the conductor cross section therefore are more suitable than standard, analytical stability models and yield a first key for successful analysis of the stability problem.

While superconductor temperature, in rather exceptional cases its local temperature distribution, can be measured with sensors that are mechanically or radiatively connected to filaments or to thin films, analytic or, preferentially, numerical simulations of superconductor temperature yield *local* temperature $T(x,y,t)$ with definitively higher resolution. The $T(x,y,t)$ are obtained from solutions of Fourier's differential equation using the solid thermal diffusivity of the superconductor material. Analytic and numerical simulations of the stability problem thus are based on the "*phonon* aspect" of heat transfer since the phonon contribution to thermal diffusivity constitutes its overwhelming part. But also radiation contributions may become important, see below.

Standard stability analysis considers the decrease, $dJ_{Crit}[T(x,y,t > t_0)]/dt$, of critical current density during a warm-up period in parallel to the increase $dT(x,y,t)/dt$ of the local "phonon" temperature, t, of the superconductor. But at very low temperature, the superconducting *electron* system (responsible for non-zero value of $J_{Crit}$) is largely decoupled from the lattice. The electron system, at least temporarily, reflects its own dynamic response to disturbances. The question is whether decay or generation of electron pairs, the "*electron* aspect" manifested by $dJ_{Crit}[T(x,y,t > t_0)]/dt \neq 0$, proceeds on another timescale, t', during decay and subsequent recombination of excited electron states to a new dynamic equilibrium. It is not clear that the timescale, t', should be identical with the traditional (phonon-based) timescale, t.

A shift, $\Delta t_{Ph/El}$, between both time scales and its impact on critical current density and stability functions has been investigated recently [4]. The shift is not constant in the conductor cross section or over its length. It can be rather large, between 1 ms and 5s in NbTi, but is very small in YBaCuO. Non-zero values of the shift lead to the more general problem of unique identification of time scales, see Sects. 7 to 8 of the present paper.

## 2    Modelling Heat Transfer in Superconductors

Radiative contributions to heat transfer in bulk solid material are vanishingly small. In thin films, however, the situation may be different. This aspect will be investigated in the present paper. This is important for conductor stability since in a superconductor, near its phase transition, even tiny temperature fluctuations arising under any heat transfer mechanism, including radiative, can locally drive the superconductor into the normal conducting state. This means all existing heat transfer mechanisms ("channels") have to be taken into account in temperature field and, accordingly, stability calculations.

Stability calculations apply the stability function, $\Phi(t)$, to calculate maximum, zero-loss transport current, $I_{Max}(t)$,

$$0 \leq \Phi(t) = 1 - \int J_{Crit}[T(x,y,t), B(x,y,t)] \, dA / \int J_{Crit}[T(x,y,t_0), B(x,y,t_0)] \, dA \leq 1$$
$$I_{Max}(t) = I(t = t_0) [1 - \Phi(t)] \quad (1a,b)$$

The integration is to be taken over the total conductor cross section, A, and the impact of local magnetic flux density, B, on critical current density has to be taken into account. An interesting example is given in Figure 13 in Appendix 3).

In Eq. (1a), the stability function applies local ratios of critical current densities, $J_{Crit}[T(x,y,t)]/J_{Crit}[T(x,y,t=t_0)=77K]$. Because of the strongly non-linear dependence of critical current density on temperature, this procedure essentially implies comparison of local temperatures, $T(x,y,t)$ vs. $T(x,y,t=t_0)$. Results for prediction of the onset of a quench, as a consequence, rely on how exact the sharp condition, $\Phi(t) = 1$ for $T \geq T_{Crit}$, can be verified, which means it relies on the uniqueness by which critical temperature can be defined. This question, as will be shown in the next Sections, is of central importance for stability predictions and for understanding of superconductor parameters in general.

Non-transparency enormously simplifies numerical analysis of the radiative transfer problem. Non-transparency is expected for optical thickness of at least τ = E D = 15, with E the extinction coefficient and D the sample thickness. The τ ≥ 15 criterion results from experience. It concerns *direct* transmission of radiation under Beer's transmission law.

However, the author several times has been confronted with this argument: "If a sample is non-transparent, there is, within the sample, no radiation at all."

This is not correct. Multi-component heat transfer, under any optical thickness, τ > 0, with solid conduction heat transfer in parallel to radiative and to possibly other transport mechanisms (channels), is the background for existence of non-zero, local, e. g. Black Body, mid-IR radiation, within a sample:

We have (a) local exchange (transformation) of energy between two parallel transport channels, because of coupling between phononic and photonic excitations and corresponding heat transfer. This applies to samples of any optical thickness.

We have (b) to take into account conservation of energy. As an example, let us return to results reported by Viskanta (1965) for coupled, stationary conductive/lradiative heat transfer in an optically thin medium (τ ≤ 1, see Figure 11 in Appendix 3. The small optical thickness is chosen just for improving clarity; qualitatively, the curves would be the same, but flattened in case the optical thickness is large. In this Figure, a stationary case is investigated, and conductive (given by the dimensionless temperature gradient, -dΘ/dτ, with thermal conductivity a constant), and radiative (Φ) heat flow are complementary to each other: At τ = 0.2, the maximum of Φ coincides with the minimum of -dΘ/dτ (only the sum of both components is constant, by conservation of energy). The main result is: Radiative heat flow intensity is not zero within the sample.

Qualitatively the same result: Heat transfer components depending on position, $q_{Cond}$ and $q_{Rad}$ adding up to a constant in stationary case and under the given thermal and optical parameter,) is obtained with samples of large optical thickness.

If a radiation beam hits a sample surface, the condition τ > 15 can easily be fulfilled with sufficiently large sample thickness. At distances, **x**, close to the surface, i. e. at a *local* τ << 1, while *total* optical thickness of the (non-transparent) sample can arbitrarily be large, the material up to this position may be transparent. Trivially, if **x** increases and finally equals total thickness, τ successively increases to large values to get the sample non-transparent. Residual radiation intensity at the rear sample surface under this condition is very weak but is uniformly distributed (Lambert's cosine law) because of a multiplicity, N = τ, of absorption/remission and scattering interactions between photons and sample constituents. The residual radiation intensity then "forgets" where it comes from and how it was created, e. g. as incident, directional or as distributed radiation.

Non-transparency thus relies on large sample thickness and on strongly limited values of the mean free path lengths, $l_{Rad}$ = 1/E, of photons. Otherwise the distribution of the weak residual radiation at rear sample surface would not be isotropic. If it is not isotropic, the sample is at least partly transparent

As a consequence, virtual observers positioned within a non-transparent material, at τ >> 1, experience radiation from closest distances only, which is a situation like in a Markov chain when the "chain" is interpreted as a radiation beam. The quasi-Markov chain is approached the better the larger the distance from the original radiation source.

Multi-component heat transfer gets solution of the combined solid conduction/radiative heat transfer very difficult, but non-transparency allows to describe both solid conduction and radiation heat transfer as diffusion processes. If applicable, this is an enormous simplification since it allows application of the so called Additive Approximation of conductivities: Total conductivity, under *this* very condition, is simply the algebraic sum of its components. This is not trivial (yet, neglecting this restriction, it has frequently been assumed in the literature). Applicability of this approximation has been investigated recently [8, 9], again by numerical simulations using very simple or more complicated cases, just to convince the readership.

Besides numerical simulations, non-transparency of the superconductor thus provides a second key to successfully simulate stability of superconductors. Strictly speaking, check of the condition τ ≥ 15 has to be performed at all relevant mid-IR wavelengths (mid-IR in case of High Temperature Superconductors). But it has been shown [6] that extinction coefficients and Albedo, all obtained from spectral values of the refractive index, did not very strongly depend on wavelength, compare Table 1 in Appendix 3 to the present paper.

Risks arising during application of standard stability models have already been explained in our previous papers [4 - 7], this shall not be repeated here. Numerical simulations instead would provide a definite answer for analysis of the stability problem. But thorough analysis of the physics behind superconductor stability imposes, even under non-transparency, at least three additional problems.

## 3   Additional Problems

In the following, we shall describe

(i) Conflicts arising from relaxation time of the superconductor electron system, after a disturbance, in relation the integration time intervals in analytic and numerical simulations,

(ii) Uncertainties resulting from strongly different propagation speed of heat transfer components,

(iii) Temporal localisabilty of events, like a quench that may question the results of standard and of numerical stability calculations.

The paper is organised following these three items.

(i) After a disturbance, how long does it take the electron system of the superconductor to arrive at a new thermodynamic equilibrium? When any physical system reaches this equilibrium (if there is any), no more excursion with time of its degrees of freedom will be observed (apart from ubiquitous statistical fluctuations). The question to be answered is not whether this equilibrium exists at all, or how long it remains undisturbed, but *how long* it takes the superconductor electron system *to* actually *reach* this state, by relaxation after a disturbance. In the superconductor literature, there is no answer that would be suitable for solution of this problem.

A general method to calculate lifetimes is provided by elements of perturbation theory, but perturbation theory breaks down near phase transitions. Time-dependent Ginzburg-Landau theory of the order parameter offers an alternative, but it would require an enormous amount of computational effort and can hardly be realised in case of transient temperatures (and, trivially, in complex superconductor geometry, like in multi-filament materials).

The relaxation time is needed to re-organise, after a disturbance, (α) normal conducting electron states (single electrons, in the presence of residual electron pairs) and (β) the un-decayed superconducting pairs, to a new thermodynamic equilibrium. Derivation and structure of the new wave functions describing reorganisation of the electron system to the new equilibrium of course would be of most interest, but solely the temporal aspect of relaxation is in the foreground in this paper.

As an alternative to perturbation and Ginzburg-Landau theories, this simulation can numerically be performed by dividing the relaxation problem into

(1) a "microscopic" part, the proper re-ordering of disturbed quantum to statistically stable, new equilibrium states, and (2) a parallel, "macroscopic" part by simulation of transient temperature from solutions of Fourier's differential equation that drives the calculations in part (1).

In the microscopic part (1), completion of each single recombination event needs an elementary time interval, $\vartheta t$. Counting the very large number of recombination events (summation over all intervals) taking into account all residual and progressively restored electron pairs, allows to obtain the total relaxation time, $\tau_{El}$. The method is explained in detail in a "microscopic stability model" [4].

In the microscopic stability model (1), let under thermodynamic equilibrium the total wave function, $\psi(t_0)$, or a set of individual wave functions, $\varphi_i(t_0)$, of which $\psi(t_0)$ is composed, describe the superconducting quantum state, at an original time, $t_0$. A thermal disturbance then requires the *whole* set of the $\varphi_i(t_0)$, not only part of them, to progressively be re-arranged (re-ordered) to a new total wave function, $\psi(t_1)$. The wave function $\psi(t_1)$ subsequently has to be anti-symmetrized, under quantum-mechanical selection rules, to the new equilibrium wave function, $\psi(t_2)$. Time needed for physical realisation of the sequence $\psi(t_0) \rightarrow \psi(t_1) \rightarrow \psi(t_2)$, yields the difference, $t_2 - t_1$, as the relaxation time, $\tau_{El}$, of the total electron system, from the disturbed system ($t_1$) to the final, new thermodynamic equilibrium state ($t_2$).

This model does not insist on a specific type of coupling in a BSC or any other superconductor. Instead of phonons that correlate two electrons to a pair, the process of coupling in the superconductor might as well be of excitonic or plasmonic or magnonic, electron-electron type. In any case, we have to divide the very small, but finite duration, $\vartheta t$, into a first step "selection" (according to selection rules) of the two electrons that preceeds a second step, the proper "coupling" of the two selected electrons, once their "selection" (to become a pair) is accomplished.

Duration of the steps "Selection" and "Coupling" in this simulation are estimated from formal analogues: (a) Calculation of "coefficients of fractional parentage" in atomic and nuclear physics; this gives the number of states over which lengths of the time steps have to be summed up, (b) a "time of flight"-concept with a mediating Boson; in nuclear physics, the Yukawa-model, as one of possible analogues, a pion, π, mediates coupling (binding) of two nucleons, and (c) the uncertainty principle.

Relaxation of decayed, single electrons to electron pairs cannot occur simultaneously with other, identical relaxation processes. This is because of the Fractional Parentage principle: If two electrons, i and j, are coupled to an electron pair, this does not mean this pair would be composed for all times by uniquely the same electrons. Since the whole electron system is described by one (and only one) total wave function, the individual electrons i and j can be replaced by individuals k and l, because of the energy uncertainty relation (electrons being spin 1/2 particles, i. e. Fermions, are not distinguishable). Therefore (and this is the Fractional Parentage principle), each recombination event N requests re-ordering of the *whole* electron system, again to a total, unique wave function. This is numerically a quasi-step by step process, which means a recombination event N can be initiated, and two electrons recombined to a pair, not before event N-1 is completed. Otherwise the Pauli principle cannot be fulfilled.

The more the system approaches critical temperature (from below, under a disturbance like absorption of radiation), the larger is the number of terms over which summations have to be performed (because of the increasingly large the number of decayed electron pairs, Figure 1a). The continuously increasing number of electron states, or of individual wave functions, $\varphi_i(t)$, therefore is responsible for divergence of $\tau_{El}$ when $T \rightarrow T_{Crit}$, Figure 1b.

As a consequence, length of numerical integration time steps, Δt, in the parallel "macroscopic" part (2) of the simulations (solutions of Fourier' differential equation) has to be reduced the more, the closer the simulation approaches critical temperature. It becomes increasingly difficult to achieve convergence of the solutions, and no convergence of the solutions finally can be obtained at temperatures very close to $T_{Crit}$. The strongly reduced Δt then collide with increasing relaxation time (the sum taken over all intervals, ϑt), Figure 1b.

Divergence of $τ_{El}$ to an apparently infinitely large value $t_∞$ near $T_{Crit}$ questions safe identification of superconductor (electron) temperature at *finite* times, t, in particular the critical temperature. We will later come back to this aspect.

Though the model [4] inherently assumes existence of a density matrix, it does not explicitly consider reordering of wave functions, φ(t), of single electron states to wave functions of electron pairs, and it does not explicitly reflect the behaviour with time of ψ(t), the temporal state of the whole electron system. The model presently does not go beyond addition of time steps. It is not clear whether a more stringent approach could be realized by considering the development with time of all initial, individual φ(t), of Hamiltonians and of the integral ψ(t).

Item (i) of the above meanwhile requests a two-fold update against [4]: (a) Finding a solution of the completeness problem of radiative transfer by means of an operator scheme, to separate fast from slow thermal transport mechanisms; this concerns item (ii) of the above, (b) A recently published paper authored by Castro-Ruiz et al. [10] on temporal localisability in quantum systems under relativity principles might shed light on temporal localisability of events also in superconductors.

When going to Item (ii), standard radiative transfer theory is incomplete because it does not differentiate between particular transition times arising under multi-component heat transfer. Calculation of transition time of a thermal excitation, under solid conduction in parallel to radiative heat transfer, has recently been reported in [7] for the YBaCuO 123 thin film superconductor. Time lag between different heat transfer components (see Figure 7a) is substantial, compare Sect. 7.

A solution suggested in [7] of the in-completeness problem divides length of a *total* time interval into appropriately chosen sub-intervals. These are collected in a matrix wherein, line by line, the different heat transfer mechanisms, according to their propagation speed (and other criteria), are separated. Division into intervals not only separates "fast" (radiative) from "slow" (solid conduction) transport phenomena in multi-component heat transfer, it separates also transition times, *within* radiative transfer, of scattering from those of absorption/remission. Also their propagation speed is strongly different.

This solution is in the present paper extended to an operator solution scheme, which simply is built up on the previously reported matrix of radiative transfer and energy equations but is more flexible and practical, and it adds more clarity to the procedure. Detailed description and an example to confirm its applicability are presented in Sect. 6.

Finally, we have a three-fold problem with time scales, item (iii) of the above: The first originates from divergence of the relaxation time, as mentioned in item (i). Also the second problem is trivial: It results from the time lag, item (ii).

But the *third* problem manifests itself in a cloud of "images" that arise from "events". Examples for events are local temperature variations at a position, **x,** that initiate or modify emission of thermal radiation (here, mid-IR photons). The radiation, when absorbed at **x**, and remitted or scattered to other positions, **x',** induce corresponding, local temperature variations: "images" (Figure 2a,b).

The point is: Any single, isolated event, by different heat transfer mechanisms (transport channels), and each transport channel individually running against a variety of obstacles produce a very large (in principle an infinitely large) number of images. Obstacles result from e. g. contact resistances, materials inhomogeneities, voids, scattering centres, particle geometry or variations of the indices of refraction. For the same event, images are not identical. Unlike the time lag between *different* transport channels, item (ii), these uncertainties, manifested by the large number of images, arise *within* each channel.

While this uncertainty may seriously deteriorate temporal localisability of events, a yet more difficult question in general concerns existence of uniquely defined, physical time scales. This is a parallel to the above mentioned paper by Castro-Ruiz et al. [10]. Missing temporal localisability in both cases might be explained from a common origin, namely from a generalised non-transparency concept (the idea is explained in Sect. 8).

The uncertainties arising from items (i) to (iii) never have been investigated in connection with superconductivity, or they have simply been assumed as negligible, which has to be questioned.

Before we in Sects. 6 to 8 in more detail discuss items (ii) and (iii), in particular the completeness problem in radiative transfer, a hypothesis shall be raised in Sect. 3.1: It is well known that electrical conductivity of superconductor materials at $T > T_{Crit}$ is larger than the expected, normal conduction value. Against traditional explanation of this phenomenon, the increased conductivity instead might be correlated with the superconductor order parameter. The extent of this contribution is unknown, but an attempt is made in Sect. 3.1 to quantify this contribution.

If this hypothesis could be confirmed experimentally, validity of the microscopic stability model [4] would strongly be supported, and the results of this model (Figure 1a,b) then also provide a key for verification of predictions reported in the next Sections.

### 3.1 A Hypothesis concerning Electrical Conductivity at $T > T_{Crit}$

Increased electrical conductivity of homogeneous metallic superconductors at $T > T_{Crit}$, in relation to the normal conducting value, usually is understood as originating, as an intrinsic property, from a "sort of fluctuating superconductivity", see Glover III [12]. This reference explains the increased conductivity as originating from superconducting carriers of finite lifetime when they are formed in the homogeneous superconductor material. Glover III claims evidence that observed rounding of the transition curve in thin samples with short electron mean free paths relies on this effect and cannot be explained from sample materials inhomogeneity.

It is tempting to alternatively explain the increased conductivity, or at least try to estimate a contribution to this effect, from two observations: (a) non-uniformity of transient temperature distributions in superconductors and large temperature gradients, (b) finite relaxation time, both items as reported in [7] and in our previous papers.

During a disturbance (warm-up of the superconductor), excursion with temperature of the phonon system is ahead the electron system, by a shift $\Delta t_{ElPh}$. The shift is a function of the (weak) coupling between both systems, it depends on temperature, and it has been calculated in [4] for the superconductors NbTi and YBaCuO 123, see Figures 13a,b to 14 of this reference. For $T(t_{Ph}) > T_{Crit}$, the relation $T(t_{El}) < T_{Crit}$ can be fulfilled provided relaxation time of the electron system, $\tau_{El}$, is large enough.

Excursion with time of superconductor transport properties in its two (normal conducting or superconducting) electronic phases then would be correlated over an "obstacle" that does not rely on materials inhomogeneity but is set by critical temperature and temperature gradients. Large temperature gradients usually result from materials inhomogeneity (different materials, different interfaces) but may be induced also in homogeneous materials by strongly temperature-dependent materials properties. These might constitute a mixture of regions of totally different electrical transport properties, like in a two-fluid model.

Under this proviso, the contribution of superconducting charge carriers, after complete relaxation, to the overall, effective electrical conductivity could tentatively be estimated from a suitable cell model if the "porosity", π (describing the dominating, normal conducting contribution to the total electrical conductivity), or the part 1 - π (of superconducting "inclusions"), would be given. Since 1 - π is not known (but certainly is very small), we will provisionally treat it as a free variable.

It is not clear that the porosity π would be constant, independent of temperature. Not only zero resistance of the superconducting state but also the residual number of electron pairs has to be accounted for in the effective conductivity. The contribution 1 - π reflects the order parameter. Since it strongly depends on temperature, the fraction 1 - π cannot be constant (since the order parameter itself is not constant).

The Russell cell model is a standard means to calculate materials and transport properties of two-component systems. The model is easy to handle: In its original version, it just contains porosity and resistivity of both, solid and porous phases, either for electrical or thermal transport, and is flexible (the role of particles and voids without much effort can be interchanged). For its description and an application see [13].
The model can be used to calculate the effective electrical resistance, $R_{eff}$, of the thermal fluctuations state by assuming it is not an intrinsic effect but the consequence of a "two fluid" mixture of normal conducting and superconducting states.

To apply the model in numerical calculations, the residual electron pair contribution has to be assigned a pseudo, finite non-zero resistance (trivially by many orders of magnitude smaller than the resistance of the normal conducting state). Predictions of the model may diverge at very small values of the porosity (this is a weak point of all cell models).

We have applied the model to the superconductor Pb, with $T_{Crit}$ = 7.2 K. Assuming π below $10^{-6}$ at temperature exceeding but very close to $T_{Crit}$ and π = 1 at T = 7.3 K (zero contribution to electrical conductivity by electron pairs), and with calculation of order parameter and relaxation time as done previously [4]. The result for the effective, specific electrical resistivity, $\rho_{El}$, is shown in Figure 1c. It is in the order of $10^{-8}$ Ohm cm and decays during cool-down of the system to $T_{Crit}$, and the specific electrical conductivity increases the closer the system approaches $T_{Crit}$ from temperatures above.

Qualitatively, the course of the curve in Figure 1c appears acceptable but quantitative agreement still has to be found. If it exists, the method to calculate the order parameter in [4] in turn would be confirmed (Figure 1c thus is coupled to the results shown in Figure 1a,b). Certainly, this needs more investigations.

The result shown in Figure 1c is at least not in conflict with standard theory, these are not questioned in general. Figure 1c just might provide a contribution to the explanation of the thermal fluctuations phenomenon.

## 4      Superconductor Samples used in the Stability Calculations

We have numerically investigated the stability problem in two $LN_2$-cooled superconductors: In the Powder in Tube (PiT) manufacturing process, tapes are prepared as first generation (1G) BSCCO 2223 *multi-filamentary* superconductors. The tapes consist of a large number of flat filaments of superconductor material embedded in a metallic (Ag) matrix. Each filament consists of thin, flat plate-like superconductor grains. Compare Figure 10a in Appendix 2; for more detailed descriptions see e.

g. Figure 1a,b of [5], or Figures 1a,b and 5 of [6], or the microscopic sections in Figure 6a,b of the same reference.

The cross section of the other superconductor, the "second generation" (2G) coated, YBaCuO 123 thin film, is described in Figure 1 of [11]. It shows a flat coil of in total 100 turns (of which in this reference the outermost five windings are numerically modelled). In the *radiative* propagation, not in the proper *materials* sense, a microscopic, mid-IR *optical* grain structure is identified even in this thin superconductor film that induces obstacles (at least by scattering) against radiative transfer, see Figure 7a-c of [6].

A flux flow resistivity, continuum cell model has been suggested recently [13]. It improves standard approximations and provides a third key for successful analysis of the stability problem and the predictability of quench in numerical simulations. This cell model improves the analyses of the stability problem at situations close to $T_{Crit}$ and in view of the related questions, items (i) to (iii) of Sect. 3.

## 5      Results: Transient Temperature Distributions

Simulated temperature distributions within the "first generation", (1G) BSCCO 2223 multi-filamentary superconductor are highly non-uniform, not only in the overall Ag/superconductor matrix (this is trivial) but also within the filaments of the multi-filamentary superconductor. In the Long Island Superconductor Cable, filament thickness is about 20 µm, thickness of a tape consisting of 91 filaments imbedded in the Ag-matrix is 264 µm. Cross section and results are shown in Appendix 2 of the present paper; this is Figure 2a of [8]), and the results are from [5] and [6].

Non-uniformity of the temperature distribution is observed also in the cross section of the (2G) coated, YBaCuO 123 thin film, see Figure 5c in [11]. Film thickness in the calculations is D = 2 µm (D taken provisionally, see Subsect. 6.3.2 and Caption to Figure 6b).

In both superconductors, the simulations indicate enormous rates, dT(x,y,t)/dt, of local conductor temperature increase when a disturbance is switched on, here a sudden increase of transport current (a fault), within 2.5 ms to a multiple of 20 of its nominal value. In the multi-filamentary BSCCO superconductor, the dT(x,y,t)/dt increase from about $3 \cdot 10^3$ K/s at the beginning of the disturbance to more than $10^8$ K/s when critical temperature of this superconductor (108 K) is exceeded, see Figure 10b,c in Appendix 2: Comparison of the results obtained for the outermost right filaments (at the axis of symmetry) yields temperature increase between 98.6 and 118.5 K, i. e. of about 20 K within 0.3 ms, which means about $6.7 \cdot 10^5$ K/s. The rate becomes still larger if the results would be considered not between 1.5 and 2.1 ms after start but at the end of the disturbance.

Note that these very large rates are local, transient values, to be understood as point-wise, transient disturbances. It is clear Finite Element simulations, like any other simulation technique, are subject to uncertainties; they arise from the Finite Element scheme itself (selection of the type of elements, structure of the mesh, values of input parameters). But accuracy of the mesh was checked by comparison of stagnation temperature with analytical results; agreement was better than within 1 mK. For thermal and optical parameters, experimental values have been applied. Besides these obvious sources of uncertainties, enormous problems can arise with divergence in simulations near phase transitions. Calculations thus have to be repeated to improve convergence, and modified (doubled) convergence schemes should be used like by the scheme shown in Figure 9 in Appendix 1.

In any case, near the end of the simulations, t > 1 ms, conductor temperature may become quite uniform but overheating, as predicted at earlier times, might already have caused damage to the conductor."

If in the thin film, YBaCuO superconductor the ratio, $I_{Transp}/I_{Crit}$, of transport to critical current is limited to 0.95 and the statistical variation, $d_{JCrit}$ of $J_{Crit}$ (an uncertainty resulting e. g. from conductor manufacture) is within only one percent, no losses from current transport in the superconductor thin film will be observed. Conductor temperature, under *this* condition is very uniform, see Figure 4a,b of [11]). But the situation changes significantly when the ratio $I_{Transp}/I_{Crit}$ closely approaches 1; see Figures 5c, 6 and 7 of the same reference.

Accordingly, in both BSCCO and YBCO superconductors, quench does not start uniformly in the conductor volume even if the disturbance, in its first stages, might result from uniformly distributed flux flow losses. Its reason is manifold:

1) temperature fluctuations arise from the statistical distributions of heat sources that are generated by absorption of radiation in the Monte Carlo simulations, 2) a statistical distribution of the superconductor critical parameters has to be assumed, to account for shortcomings in conductor manufacture, handling and winding, compare the schematic Figure 12 in Appendix 3 (for example, we cannot expect strictly uniformly distributed critical current density), 3) the samples experience different heat transfer conditions, between superconductor and matrix material or substrate material (contact heat transfer) or to the coolant, at the solid/solid and solid/liquid interfaces (convection), respectively, 4) fluctuations of magnetic field (penetration depth) that have impacts on local critical current and thus on transport current density.

Also these details cannot be handled with standard stability models.

In both BSCCO and YBCO superconductors, non-uniform temperature distribution and, accordingly, *local* generation of quench is in strong contrast to the traditional assumptions.

In both superconductors, zero loss, flux flow resistive and Ohmic resistive states therefore may coexist, within finite time intervals in the conductor cross section. There is no laminar flow-like current transport; instead, transport current under disturbances percolates through the conductor. If there are no, or no longer, zero resistance transport channels, the current selects those of lowest resistance, a process that *in summa* provides minimum losses.

Like distribution of transport current, losses in the conductor cross section may be different in each time step and at each length-position of the conductor.

# 6 A Completeness Problem of Multi-component and Radiative Heat Transfer

The theory of radiative transfer is explained in standard volumes like [14] or [15]; focus of [14] is rather on technical issues while the frequently cited [15] explains classical theory of radiative transfer. Applications to single fibres, multi-filamentary superconductors and thin films has been described in our previous papers [4 - 9, 16]; details will not repeated here.

This Section refers to item (ii) of Sect. 3. The completeness problem, from the rigorous radiative transfer aspect, has been explained in Sect. 4.2 of [7]. The situation is in some aspects similar to public traffic (vehicles with strongly different speed), or when considering diffusion of different species, like with mixed gases or in soldering at solid/liquid interfaces. These and a large variety of other diffusion-related problems can be described with a set of generalized Burgers' equations. But diffusing species like cars, gas particles or alloy atoms can be distinguished clearly, which is not the case in radiative transfer: There is only one species (photons or beams thereof), and photons can be differentiated with respect to only wavelength (after remission or after elastic or inelastic scattering) and to polarisation.

In multi-component heat transfer, the situation can be specified by means of a chain of thermal resistances. The resistances are switched in parallel, $1/R_{Total} = 1/R_{SolidCond} + 1/R_{Rad}$, with the total resistance, $R_{Total}$. With a temperature difference, $\Delta T = T_1 - T_2$, over the chain, the total heat flow amounts to $Q = \Delta T/R_{Total}$.

From only the "height" (magnitude) of the resistances (obstacles to solid conduction and radiative heat flow), no conclusions can be drawn on the ratio of propagation velocity of a thermal wave (remitted photons, phonons) through the material against scattered radiation.

In multi-component heat transfer (here: solid conduction and radiation in thin films), transit times of the carriers of thermal excitations (phonons, photons) through a sample are strongly different (Figure 7a in Sect. 7). Scattered radiation propagates by the velocity of light, c/n, at any given position x < D. The photons do not "wait" for arrival of remitted radiation and of phonons. Scattered radiation has overcome its obstacles long before photons from remission and phonons arrive at the same position.

But the standard Equation of Radiative Transfer (ERT), though it includes absorbed, remitted *and* scattered radiation to yield directional intensity, i', does not explicitly contain time. Its solutions receive their time dependency only indirectly from another equation, the "Energy Equation", see below, Eq. (2b).

The ERT reads

$$di'(\tau)/d\tau = -i'(\tau) + [(1 - \Omega) i'_{BB}(\tau) + \Omega/(4\pi) \int \Psi(\omega_i,\omega,\tau) i'(\tau) d\omega] \qquad (2a)$$

with i' the directional intensity, $\tau$ the optical thickness, $d\tau = E\, ds$, E the extinction coefficient, $E = A + S$, with A and S the absorption and scattering coefficients, ds a differential of the radiation path length, $i'_{BB}$ the directional, black body (BB) intensity, $\Omega = S/E$ the Albedo of single scattering, and $\Psi$ the scattering phase function; an index to specify wavelength has been omitted in Eq. (2a). The quantities $\omega_i$ (incident radiation) and $\omega$ indicate solid angles.

The term in square brackets in Eq. (2a), [(1 - Ω) i'$_{BB}$(τ) + Ω/(4π) ∫Ψ(ω$_i$,ω,τ) i'(τ) dω], usually is called the "source function". Beer's law, di'(τ)/dτ = -i'(τ), is simply a special case of the ERT in that the source function disappears.

Without the Energy Equation, the ERT (like Beer's law), would describe stationary states.

The ERT thus operates with a *common* (one and only one) time variable for each of the radiation transport channels. This is one point that deserves attention.

Second, for conservation of energy, local temperature, T, in dependence of time, T = T(x,y,t), is obtained from only the Energy Equation (denoted as "EQ" in the following). The EQ contains (a) the non-scattering, radiative energy transport channels, and (b) the solid conduction.

The EQ reads

$$\rho\ c_p\ \partial T/\partial t = \mathbf{div}\ (\mathbf{q}_{Cond} + \mathbf{q}_{Rad}) \quad (2b)$$

In the EQ, **q**$_{Cond}$ + **q**$_{Rad}$ denote heat flux vectors due to conduction and radiation, respectively.

Both heat fluxes, **q**$_{Cond}$ and **q**$_{Rad}$ (with **q**$_{Rad}$ including solely absorbed/remitted, not scattered radiation) depend on temperature. They therefore are coupled to each other by the calculated transient temperature profiles in the object. While this concerns coupling of two heat transfer components, the solutions of the EQ, T = T(x,y,t), enter the ERT by the Black Body term (one part of the "source term" of the ERT), and this couples also the two equations, ERT and EQ, to each other.

A solution of the total (multi-component), time dependent heat transfer problem therefore is obtained only if both equations, ERT and EQ, can be solved simultaneously (an extremely difficult task, *per se,* in the standard formulation).

But the EQ also uses the same radiation intensity, i', as the ERT (when integrated over all solid angles) that in the ERT are derived *with inclusion of scattering* while scattered intensity is *not (and must not be) included* in the EQ. Scattered radiation (if it is scattered elastically) does not contribute to temperature excursion with time. This finally constitutes the "incompleteness" problem.

In summary, a theory that

- does not account for strong differences in the propagation velocities of the radiative and (solid) conductive transport mechanisms, and that
- applies scattered radiation intensity in both ERT and EQ on a single time scale

remains incomplete.

Few attempts to solve the completeness problem in rigorous radiative transfer have been reported in the (open) literature, like in Sect. 21.6 of Siegel and Howell [14] and citations therein. But only radiation propagation under absorption/remission (no other modes of heat transfer) were considered in the time dependent equation of radiative transfer.

## 6.1   A possible Solution

As a way out of the incompleteness problem, we have in [7] suggested a matrix the elements of which are of the ERT and energy equations (EQ) types. A series of ERTs and, correspondingly, a series of energy equations, are applied. The matrix shall in the following be arranged to a more practical scheme.

The series of ERTs and energy equations can be re-arranged as operator equations, with a symmetric n x n, diagonal matrix **M**, and a column vector, **V**, of n lines. The matrix contains the ERT, and the column vector provides the corresponding energy equations, EQ.

The number of matrix **M** and vector **V** elements has to be chosen in appropriately dimensioned time intervals according to the physics of the transport processes and the speed by which its components proceed through a sample. Propagation of radiation by scattering clearly is the fastest process. Specification of the length of time intervals and of the integration steps in the Finite Element procedure (solution of Fourier's differential equation) has to fulfil convergence criteria.

An example for application of the operator scheme is given in Figure 3a: It describes a superconductor thin film and its co-ordinate system that specifies area elements and, by rotation against the symmetry axis (thick solid line), volume elements. The elements are used for both Finite Element and Monte Carlo simulations. The target, a circular surface, is subject to a sudden heat load. Origin of the load is arbitrary. In the Finite

Element scheme, the load is thermalised in the conductor by solid conduction (described by the standard Fourier conduction law) and, in the Monte Carlo simulation, by radiation beams emitted from the target to the volume elements. All materials parameters depend on temperature.

In this example, just to demonstrate applicability of the method, it is sufficient that the matrix for simplicity is of only the 4 x 4 type, and the column vector accordingly contains four rows. The results not only show that a difference exists between temperature distributions applying solid conduction plus radiation (the realistic scenario) or only solid conduction (both contributions here described as conductivities); the difference is clearly seen (Figure 4a-d against Figure 5a) when comparing temperature fields and maximum conductor temperature, Subsects. 6.3.1, 2.

With these dimensions of **M** and **V**, a second difference is seen if, later, (Figure 6a,b) the "standard procedure" (solid and radiative conductivity including the Monte Carlo simulation) is compared with results obtained with the operator concept.

In all calculations, a check of the stagnation temperature obtained with the simple energy balance, and the results from the operator concept, yields almost perfect agreement.

The matrix **M** operator reads

$$\begin{pmatrix} di'(\tau)/d\tau = -i'(\tau) + \Omega/(4\pi) \int \Psi(\omega_i,\omega,\tau)\, i'(\tau)\, d\omega], \ \Omega = 1 & 0 & 0 & 0 \\ 0 & di'(\tau)/d\tau = -i'(\tau) + \Omega/(4\pi) \int \Psi(\omega_i,\omega,\tau)\, i'(\tau)\, d\omega], \ \Omega = 1 & 0 & 0 \\ 0 & 0 & di'(\tau)/d\tau = -i'(\tau) + [(1-\Omega)\, i'_{BB}(\tau) + \Omega/(4\pi) \int \Psi(\omega_i,\omega,\tau)\, i'(\tau)\, d\omega], \ \Omega < 1 & 0 \\ 0 & 0 & 0 & di'(\tau)/d\tau = -i'(\tau) + i'_{BB}(\tau),\ \Omega = 0 \end{pmatrix}$$

and the column vector **V** is given by

$$\begin{pmatrix} \rho\, c_p\, \partial T/\partial t = \mathbf{div}\,(\mathbf{q}_{elSC}),\ \Omega = 1 \\ \rho\, c_p\, \partial T/\partial t = \mathbf{div}\,(\mathbf{q}_{SolidCond+elSC}),\ \Omega = 1 \\ \rho\, c_p\, \partial T/\partial t = \mathbf{div}\,(\mathbf{q}_{SolicCond} + \mathbf{q}_{inelSC+Abs/Rem}),\ \Omega < 1 \\ \rho\, c_p\, \partial T/\partial t = \mathbf{div}\,(\mathbf{q}_{SolidCond} + \mathbf{q}_{Abs/Rem}),\ \Omega = 0 \end{pmatrix} \quad \begin{array}{l} 0 \leq t \leq t_1 \\ t_1 < t \leq t_2 \\ t_2 < t \leq t_3 \\ t_3 < t \leq t_4 \end{array}$$

The symbol $\Omega$ denotes the Albedo for single scattering. From the results obtained in [6], $\Omega$ is nearly constant. Its weak temperature dependency has been integrated into the calculation of the extinction coefficients. But this is still an approximation only, because it is not clear that $\Omega$ constant is generally applicable to also multiple scattering in dispersed, radiative structures (compare Figure 7a-c of [6]) of YBaCuO 123 materials). To simulate this uncertainty, the Monte Carlo simulations described in Sect. 7 accordingly simulate the Albedo as a statistical quantity.

Dimension of all vectors **Q** in this paper is W, and dimension of the vectors **q** (like $\mathbf{q}_{elSC}$, $\mathbf{q}_{SolidCond+elSC}$, $\mathbf{q}_{SolidCond}$, $\mathbf{q}_{inelSC+AbsRem}$) is W/m$^2$.

The $t_i$, $t_j$ (right to the column vector) define those time intervals within which elastic or inelastic scattering, absorption/remission and scattering, or only absorption/remission, or only solid conduction, or combinations of all, respectively, contribute to total heat transfer. All transport mechanisms occupy their individual time scales according to their individual transit times, Figure 7a, and by their individual strengths. The matrix can be further extended to sets that correspond to different source functions or wavelengths.

The operator product

$$\mathbf{M} \times \mathbf{V} \qquad (3)$$

yields a series of solutions (a solution, column vector) for the transient temperature distribution, T(x,y,t), during each of the time intervals. By the symbol "×" in Eq. (3), lines and columns of **M** and **V**, respectively, are coupled to specify the solution problem within the given time intervals.

In this scheme, boundary conditions (intensity, temperature) may replace the more difficult modelling of exponential decay of radiation incident into, and absorbed/remitted in, the superconductor sample.

Application of the operator equations is laborious. But if the object under study is non-transparent, modelling of radiative transfer drastically simplifies to diffusion solutions. This allows to apply the Additive Approximation (under the condition the sample is non-transparent to mid-IR radiation).

When a distribution of instantaneous, initial heat sources is provided by the Monte Carlo simulation of the absorption of a large number of bundles, this distribution is equivalent to an initial temperature distribution (Carslaw and Jaeger [17]). This theorem allows to treat the whole thermalisation problem (solid conduction in parallel to radiation) as a *conduction* process, a very strong reduction of the complexity of the method.

Non-transparency and the Carslaw and Jaeger theorem provide the ideal situation to explain incompleteness of Radiative Transfer and to justify separation of the total simulated period into time intervals by application of the operator scheme.

## 6.2   Application to a Thin Film Superconductor

It has to be shown that the scheme explained in the previous Subsection really works in numerical simulations. This is not clear: The scheme involves repeated *changes* of input variables, mostly radiative, e. g. from scattering to absorption/remission (i. e. not just considering temperature dependencies of variables) during the running, numerical integration procedure, which means convergence might seriously be disturbed.

In order to enhance the effect exerted by the anisotropic thermal conductivity in this example, with an anisotropy ratio χ (approximately χ = $\lambda_{ab}/\lambda_c$ = 10), the crystallographic c-axis in Figures 4a-d, 5a,b and 6a,b (not in Figure 6c) has been oriented parallel to the sample x-axis (this is in this study for *systematic* purposes only; this orientation would not be suitable to obtain large, zero-loss current transport).

For an over-all view of what is to be expected from the calculations, Figures 4a-d and 5a,b first show results applying the standard procedure, which means without dividing the time axis into intervals. Later, Figure 6a-c, the standard procedure is replaced by the operator concept.

## 6.3 Results obtained with standard theory and with the operator concept

### 6.3.1 Results obtained from the standard procedure

Figure 4a-d shows temperature distribution within the sample at $t = 5 \cdot 10^{-8}$ s after start of the disturbance. In Figure 4a,b, emission of the beams is not from a sharply defined, single position but is variable within the target area: It is assumed, as a realistic case, that the positions are randomly distributed but concentrated around given cluster points (like elements of a mathematical series, here the individual emission points from which beams are emanated).

A strong temperature increase would result (Figure 5a) if the excitation is only on the target surface, without generating heat sources within the sample. This is due to the small, solid thermal diffusivity of the material that induces a thermal "mirror" effect. In contrast, the Monte Carlo simulations distribute part of the total thermal load by emission to the interior of the sample (though exponentially damped) and thus serves for strongly reduced front side temperature, by conservation of energy.

Temperature of a considerable part of the conductor cross section exceeds $T_{Crit}$ (lower part of Figure 5b).

### 6.3.2 Results obtained with the Operator concept

Application of the operator concept to the same physical conditions using matrix **M**, column vector **V** in Eq. (3) yields the dark-blue diamonds in Figure 6a. Total length of the pulse is divided into eight single, short pulses to allow repeated application of the Carslaw and Jaeger theorem:

Each single pulse (and its distribution by the heat transfer mechanisms) prepares the initial temperature distribution for the next pulse.

Near the final end of the series of pulses (t = 8 ns), the difference to the standard procedure, at $T < T_{Crit}$, is very small, below 0.2 K (see the results within the black ellipse in Figure 6a). But this difference reflects only the temperature excursions at a single position (x=0,y=0). A possible influence on the stability function can be found only from the integral over the total cross section (Eq. 1a) and from mapping of the temperature field onto the field of critical current density. See a check of this expectation in a subsequent paper; in the present paper, we concentrate on primarily the applicability of the operator scheme.

If $t(n_{last})$ denotes the temperature obtained at the end of the load step $n_{last}$, the division of the time axis is for all times t given by $t(n_{last} - 1)$ + dtime,j. For all load steps (each of length 1 ns), the values of dtime,j are given as noted in the Caption to Figure 6a. In comparison to Figures 4a-d, 5a,b and 6a,b, the orientation of the c-axis in Figure 6c is parallel to the y-axis of the sample.

Within t ≤ dtime,1, the Albedo is set to $\Omega = 1$ (solely scattering) and for solid conduction heat transfer. During this period, a considerable part of the intensity emitted from the target is lost from the sample; scattering may direct the residual beams to e. g. neighbouring thin films, substrates or electrical insulation, which means it does not contribute directly to increase of temperature within the proper superconductor thin film. In all four intervals, $\Omega$ is set according to the specifications of the column vector, see above.

While in Figure 6a, the deviations between standard procedure and operator concept are very small, this changes significantly when different weight is given to radiation by variations of the dtime,j (Figure 6b). In this Figure, decreasing values of dtime,3, from 0.9 to 0.5 and 0.1, are selected according to transit times shown in Figure 7a in order to put increasingly more weight to solid conduction. Reduction of dtime,3, because of the much larger solid conductivity, accordingly results in a continuous decrease of nodal temperature, compare the diamonds in the black ellipse in Figure 6b.

The observed uncertainty of solid temperature seen around $T_{Crit}$, here in the early stages of temperature excursion, t = 10 ns after onset of the disturbance, Q, therefore questions the predicted superconductor stability. At later times, data points almost coincide but then $J_{Crit}$ is zero, in any case, and accordingly does no longer have impacts on stability against quench.

The same result would inevitably apply if film thickness is reduced to D < 2 µm because of increasing contribution by radiation to total heat transfer. In magnetic coils, thin YBaCuO 123 films usually are prepared with thickness below this value. Then care has to be taken in the stability analysis because reduced thickness D might question non-transparency and applicability of the Additive Approximation.

Short load steps in Figure 6a,b (in units of $10^{-9}$ s) shall account for e. g. incident, high energy particle radiation. Extended length of load steps ($10^{-5}$ s, Figure 6c) is applied to simulate the release and transformation of mechanical stress to thermal energy by conductor movement under magnetic (Lorentz) forces.

Figure 6a,b demonstrates that the operator concept is really applicable. Trivially, the small differences between standard and operator concept are due to the small length of the load steps. In contrast, Figure 6c is intended to (i) confirm (within the same sequence in time of heat transfer mechanisms) that the operator concept works also with increased length of the pulses and (ii) under variations of materials conductive properties. This result is not trivial; variations of thermal parameters during numerical integration can lead to serious convergence problems, and the variations become increasingly important if length of pulses and time intervals is expanded.

After t = 8 ns (Figure 6a,b) and $8 \cdot 10^{-5}$ s (Figure 6c), respectively, the calculation is continued with solid conduction plus radiation, like in the standard procedure or conventionally by simple solid conduction. Pure radiative signals from the incident pulses have died out at the end of each load-step.

If in turn temperature distributions are given, variations of the dtime,j according to Figure 6a-c may be suitable to extract, from approximations to given temperature excursion, the temporal sequence of heat transfer mechanisms, by an n-dimensional $\chi^2$-fits of calculated to experimental data (in the present case, n = 4, per load-step). Measurements have to be taken at least at n positions in the cross section.

The whole procedure again is laborious, but it is an extension of traditional laser-flash experiments: The operator concept yields a time sequence of thermal diffusivity that resolves the contributions from different heat transfer mechanisms. Traditional interpretation of laser flash experiments does not provide this information.

The suggested operator scheme allows solution of the combined solid conduction parallel to radiative transfer also in increasingly complicated modelling problems, e. g. if a radiative disturbance is emitted from a source outside the proper superconductor, like a heat pulse from a normal conducting layer that, under current sharing, as a stabiliser contacts and mechanically stabilises the superconductor thin film. Explicit solution of the Equation of Radiative Transfer then would not be possible.

A very similar problem arises in the well-known solution problem experienced in the Parker and Jenkins method to remotely determine the thermal diffusivity of thin films (see the explanations given in Sect. 4 of [7]). Besides the laser-flash measurement, another similar situation arises in several "simple" engineering applications like heat conduction in packed beds of particles by conduction (short ranged and comparatively slow) and radiation through the voids of the packed bed (long ranged, and with very small transit time of the scattering contributions).

## 7    Time Lag from Monte Carlo Simulations

The average radiative *temporal* length, the effective transition time, $t_{Trans}$, of beams under purely scattering and direct transmission, which means the temporal dimension of a corresponding "cloud" of images, is in the order of only $10^{-14}$ s, with a standard deviation of between 1 to 2 x $10^{-14}$ s, see Figure 10a,b of [7]. Direct transmission means: The scattered, residual intensity that remains after each interaction of the beam with an "obstacle". It is detected at the rear surface of the 2 µm, YBaCuO 123 thin film.

The number, N, of radiative/solid interactions amounts to about 30 to 60 (black diamonds in Figure 8 of [7]), for a mean value obtained with $5 \cdot 10^4$ bundles. It substantially exceeds the critical (direct transmission) optical thickness, $\tau = 15$, because only few of the radiation paths are strictly parallel to the surface normal; the large majority follows zig-zag paths. The large number, N, accordingly confirms applicability of the non-transparency approach in the present samples.

But the *real* situation (heat transfer, not only by solely radiation) involves *combined* solid conduction and radiation by scattering and absorption/remission. Each bundle after absorption at an obstacle would be split into an arbitrarily large number of remitted, "conductive" bundles that is too complicated to be simulated with solely a Monte Carlo method. We yet can treat this realistic but difficult, total thermal transport problem by application of an approximate diffusion solution of Fourier's differential equation, see standard textbooks on Heat Transfer, e. g. [18], Eq. 4.3-26, $L = C (a_{Th} t)^{0.5}$. An extension to also radiative transfer can be applied: With $a_{Total} = a_{SC} + a_{Rad}$ the total diffusivity, the transition time is estimated from

$$t_{Trans} = (L/C)^2/(a_{SC}+a_{Rad}) \tag{4}$$

with L = sample thickness, C a constant that is obtained from the standard, "1 percent" condition; C = 3.6 in a flat sample.

But this is only an approximation to the present problem: Eq. (4) assumes that surface temperature of a flat sample suddenly jumps to a constant, stable value. Nevertheless, Eq. (4) has successfully been applied in the literature to also other disturbances (except for very short pulses that require full solution of Fourier's differential equation, like in the series expansion reported in [16]).

Application of Eq. (4) works if the Albedo $\Omega < 1$ and if the diffusivity $a_{Rad}$ is restricted to solely the absorption/remission part of the radiation.

The summation $a_{Total} = a_{SC} + a_{Rad}$ is again justified by the Additive Approximation since the diffusivities contain the corresponding thermal conductivities. The other values in the diffusivity are approximately constant specific heat and density.

The results (Figure 7a) indicate existence of an enormous time span between the maximum (purely absorbed/remitted radiation, a fictitious case) and the physically realistic minimum (solid conduction plus radiation; pure scattering again is a fictitious situation). Here we note that there is no superconductor the radiation extinction properties of which would rely solely on absorption/remission ($\Omega = 0$). The transmission time for the combined solid conduction and radiative (absorption/remission) heat transfer estimated from Eq. (4) thus delivers an upper limit, $t_{Trans} \leq 5 \cdot 10^{-2}$ s, at T close to $T_{Crit}$. This is rather independent of the Albedo, $\Omega$.

The uncertainty interval, $dT(x,y,t)$, of the temperature, at $x = D$ (D the thickness of the thin film), then is calculated using temperature variations, $dT(x,y,t)/dt$, multiplied by the transit time, $t_{Trans}$. The (large) variations $dT/dt$ are obtained from the simulated, transient conductor temperatures.

The final result, $dT(x,y,t)$ (Figure 7b) is explained by the transit time of $5 \cdot 10^{-2}$ s transformed to a temperature uncertainty arising within this period of about 0.7 K. While this is small, dT(x,y,t) = 0.7 K is already larger than uncertainties observed in standard, critical temperature measurements.

This result is well comprehensible from the time lag *between* different heat transport channels and from statistics *within* a particular channel. But according to item (iii) in Sect. 3, the situation is still more complex. The question, whether or not uniquely defined, local physical time scales exist, item (iii) of Sect. 3, is not just academic.

## 8    Existence of uniquely defined Time Scales

First conclusions from the previous Sections can be drawn:

(i) A considerable part of electron pairs, during a disturbance and its relaxation, remains un-decayed (strictly speaking: is continuously restructured, by keeping the number of constituents statistically constant) and serves for zero-loss current transport.

During relaxation, electron pairs and single electrons, continuously generated under a disturbance, temporarily co-exist, like in classical, two-phase systems. This is the background of the attempt reported in Subsect. 3.1 to suggest an alternative explanation of the additional electrical conductivity at $T > T_{Crit}$. This observation might explain also the standard R(T), resistance vs. temperature diagrams, when closely inspecting on the T-scale the steep increase near $T_{Crit}$ from zero to non-zero resistivity.

(ii) As long as relaxation at each temperature is not completed, which means, as long as no thermodynamic equilibrium at these temperatures is attained, time cannot uniquely be defined: Time has to be understood as being strictly correlated with the thermodynamic "background" of electron states

.

Time $t_{El}$ is to be understood as a variable defined for solely the electron system. As mentioned, there is a time shift $\Delta t_{Ph/El}$ between $t_{El}$ and its phonon counterpart of lattice excitations, $t_{Ph}$. There are situations where $t_{Ph}$ is ahead $t_{El}$ or vice-versa but never is $t_{Ph} = t_{El}$ during disturbances.

(iii) Relaxation proceeds fast, but it is a continuous process stepping forward in a very large number of small elementary relaxation processes; it is not a sudden event. Time $t_{El}$ is correlated with each of these elementary processes. This has serious consequences to explain critical temperature, $T_{Crit}$:

If $T_{Crit}$ exists, it can be understood as being either

- a rough description of a thermodynamic *non-equilibrium electron state* (its standard but only approximate interpretation), which means from purely thermodynamic standpoints, $T_{Crit}$ cannot be explained as a thermodynamic temperature at all
- or $T_{Crit}$ correlates, as a real thermodynamic *equilibrium variable*, with a corresponding, really existing equilibrium electron state; unfortunately, this electron state cannot be reached within finite process time (electron temperature may increase to any high value, but corresponding electron states do not converge to a final equilibrium value).

This is because any (superconductor electron) temperature, even if it is infinitely close to $T_{Crit}$, cannot defined as a thermodynamic variable as long as no thermodynamic equilibrium of the electron system is attained. In any standard warming-up experiment, the approach to the electron temperature $T_{Crit}$ by phonon temperature increase from temperatures below, too strongly increases electron relaxation time so that the electron equilibrium state finally (at $T_{Crit}$) becomes out of reach within reasonable experimental or computational (process) periods.

Accordingly, $T_{Crit}$ either does not exist at all, or its background, an equilibrium electron state, cannot be attained within finite time (Figure 8 of [11]).

Critical temperature also cannot be understood as the *limes* of a series of (simulated) temperatures, T, each correlated with a corresponding electron state. These states (even if they all might be equilibrium states) do not converge within finite time to the state corresponding to $T_{Crit}$.

A uniquely defined $T_{Crit}$ thus cannot exist and is not available for sharp, definite decisions on superconductor resistance states and stability. Since each intermediate state (equilibrium obtained at temperatures T < $T_{Crit}$) correlates with process time, and since the series of equilibrium states does not converge within finite periods, it is also time, $t_{El}$, that does not converge to a value definitely identified as $t_{El} = t_{El}(T_{Crit})$ as the limes of a parallel series.

A usual way out of the problem is to understand $T_{Crit}$ as a strongly *non-equilibrium* electron quantity (when zero-loss current transport breaks down) but simply continue with the approximate interpretation that it is the phonon temperature, and accordingly, $t_{Ph}$, that are in parallel detected in the experiments. But this view does not solve the problem of uniqueness of physical time.

Does this conclusion ($T_{Crit}$ a strongly *non-equilibrium* quantity) apply to also critical current density, $J_{Crit}$, and thus confirm the microscopic stability model? Zero-loss current transport is observed as long as apparent, thermodynamic equilibrium states are obtained (the yellow diamonds in Figure 1a). At $T_{Crit}$, $J_{Crit}$ is zero. Critical current density, like critical temperature, thus has to be understood as a strongly non-equilibrium quantity.

## 8.1 Localisability of Events and Images on Time Scales

The following in some aspects appears to be parallel to a very recently published paper by Castro-Ruiz et al [10]: The authors investigate temporal localisability in quantum systems under relativity principles. The authors explain: When clocks interact gravitationally, the temporal localisability of events becomes relative, which indicates a signature of an indefinite metric.

Let events, 1 and 2, at different positions be given, under the observation of three experimenters, A, B and C. Then A might report the events occur simultaneously, while B reports 1 occurs before 2 and C recognises 2 before 1. The reports provided by A, B and C depend on their reference systems: A universal time scale for the three observers that uniquely books events 1 and 2 in a common logical order on a time scale does not exist. From relativity, there is also no universally defined "instant". This is the classical result.

But the paper by Castro-Ruiz et al [10] goes forward: They consider a set of multiple clocks, with description of the evolution of a quantum system in which the space-time metric is indefinite, due to gravitation in these systems by superposition of energy or position eigenstates. Obviously, the authors consider existence and inter-dependencies of multiple time scales in transparent space (but transparency of their space is really the question). On the other hand, the Schrödinger equation is form-invariant under transformation from one time reference frame to another. But when checked by items (a) to (e) and their extensions, see these later in Sect. 8.2, the situation would be better explained by a more general notion of non-transparency.

The present paper generally questions localisability if a system is non-transparent, either in classicial or in relativistic states. In terms of a logical order that could be observed by the three experimenters A, B and C, unique localisability gets lost not only in view of three time scales related to relativistic conditions (namely those of A, B and C) but also for, in principle, *all* time scales if these were constructed within a non-transparent medium.

In another step forward, and in a more general conception than only in the mid-IR optical sense or between gravitationally coupled clocks (or in case of entanglement), non-transparency in a more general sense could be the common background of these three and possibly of other situations. The problem then is how to define a general and acceptable understanding of non-transparency.

## 8.2 Contrasting Transparency by Non-transpareny

Here it is helpful to recall how transparency of a solid medium like a thin film is defined (if it would exist at all) and how it opposes non-transparency in the strict optical sense.

Assume a thin film sample positioned between two parallel planes at **x** = 0 and **x** = D and let radiation beams be emitted, in direction normal to the planes, from a small target area located at **x** = 0 (the target parallel to the planes). The following items (a) to (e) apply also if emission from the source is isotropic, or if emission comes from a thermal source of infinitely small or of extended, non-zero cross section (the assumed geometry is not very decisive, one can for the following imagine quite a different setup).

If a detector responds to an original beam or to an original distribution of beams all without interferences emitted from the target and received at a position **x** ≥ D, and if the medium is transparent, the detector will be able to differentiate between

> *(a)* radiation emitted by the source at constant power or wavelength, with the radiation source at different (axial) positions of the target, or
> *(b)* radiation emitted at variable power or wavelength, but with the radiation source at a fixed, single position,
> *(c)* monochromatic radiation intensity emitted by the source at different power,
> *(d)* radiation intensity emitted at constant power but at different wavelengths,

*(e)* single isolated pulses, or series thereof, and periodic radiation sources, all emitted from any (stationary) position or at any wavelength or at any time or frequency, are received at x ≥ D or at any other co-ordinate, under conservation of the temporal order of emission and of detection.

If any of the above listed items *(a)* to *(e)* is violated, the sample is non-transparent.

Violation of item *(e)* would not get this conclusion into contradiction with the HOM experiment [19, Figure 2]: Let two photons emitted from the same source hit a beam splitter, under different directions. If they are scattered onto two separate detectors, the photons are distinguishable only if they do not arrive simultaneously at the beam splitter, a result that cannot be explained by classical theory.

If in a transmission experiment the residual angular distribution of radiation leaving the sample on its rear surface is isotropic (i. e., if it follows the cosine-law), the sample again is non-transparent (this property provides an easily applicable check to decide whether the sample is non-transparent and the Additive Approximation is justified).

Trivially, in order to successfully realise items *(a)* to *(e)* in an experiment, the time interval between any variation of wavelength, duration, intensity or position of the source must be shorter than the characteristic time of the detector. The observer accordingly will only then "almost immediately" notice any variation of the emitted signals.

Violation of items (a) to (e) and their extensions thus might provide a tool to identify a possibly existing common background of generalised non-transparency in other physical situations: Gravitationally coupled clocks and entanglement.

## 8.3 Mapping Functions

Previously, we have described transparency of the space between two stationary, flat planes at **x** = 0 and **x** = D by means of mapping functions, f[e(**s**,ζ)], that create images, e(**x**,t), of events, e(**s**,ζ); compare Figure 3 in [6], the detailed description is not repeated here.

In short, assume that at the position, **s**, a set consisting of an arbitrarily large number of (stationary) events, e(**s**,ζ), can be defined, with ζ a provisional (heuristic) "scale" at this position that simply counts the events, like emission of a radiation beam from **x** = 0 (or from any other position, **s**, within the space) according to their temporal order of emission. Provisionally, we can say the events and their sequence simply are "booked" by their order on the heuristic scale, ζ. For transfer (promotion) of this scale to a genuine, uniquely and unambiguously defined "time scale", t, the time interval between any two successively booked events, e(**s**,$ζ_1$) and e(**s**,$ζ_2$) at the same position, **s**, potentially must be infinitely small, which means the distribution of the events on this scale, ζ, must be "dense".

The property "dense" means: A distribution of elements, e(**s**,ζ), and of their images, f[e(**s**,ζ)] = e(**s**,t), that reflects the distribution of elements of the set **R** of real numbers. Between any of these, there exists an unlimited number of other elements belonging to the same set. The number of elements on the (then genuine) time scale, t, is uncountably large, yet the elements are successively ordered (like the elements of the **R** are ordered). The set **R** can be expanded to **R**$^3$, if needed.

A complete (perfect) correlation between time scales is possible only if the medium is transparent. If so, the mapping functions can be interpreted likewise as "bijective" correlations (injective and surjective), in the mathematical sense; this definition applies to also sets of an unlimited (uncountable) number of elements like **R**. Time scales, t, to be uniquely and unambiguously defined, cannot exist without dense sets of events or images.

Reversely, while conservation of the order is provided in case of *transparent* media, this is not fulfilled in a *non-transparent* medium, neither in space nor in time. The bijective correlation between events and their images gets lost in non-transparent objects.

## 8.4  Consequences

A single time scale to exist in transparent media implies existence of potentially an arbitrarily large number of other time scales, all correlated among each other. A proof of this conclusion can be attempted as described in Appendix 4.

Time scales could be fabricated also in non-transparent media, but they are not uniquely defined and cannot be correlated to each other. Totally uncorrelated time scales means: In non-transparent media, no correlated time scales exist at all, as a logical conclusion; otherwise, if a set of uncorrelated time scale existed, it is not clear which of these should be preferred.

The enormous fluctuation of $t_{Trans}$ seen in Figures 7a and 8a,b excludes any individual, unique *mathematical, logical* correlation between *events* (bundles) emitted from the target and*, physically*, macroscopic observables (the *images*), like a temperature variation, when radiative beams emitted at **x** = 0 are absorbed at **x'** ≠ **x**.

By the lemma in Appendix 4, the fluctuations in Figures 7a and 8a,b in non-transparent media also exclude more generally any correlation of *time scales* at **x** with time scales at the other co-ordinate **x**'.

## 9      Summary and Open Questions

The results reported in the previous Sections are intended to improve present understanding of time, temperature, quench and, tightly related to these items, non-transparency, if a superconductor is already close to a phase transition. Observation of images uniquely correlated to their events is the basis for successful, perfect local stability analysis (temperature variations as the answer of the superconductor to events arising at different conductor positions).

The correlation between temperature fields or their variations (events) with their images in exceptional situations might be bijective. Trivially, this is, for example, not the case with critical current density but also origin of local temperature variations under multiple component heat transfer during disturbances is not at all uniquely defined.

It is not the situation at the beginning of a local quench that is the most critical. Instead, those instants deserve the most attention when the superconductor very closely has approached critical values of its parameters.

If the superconductor is already close to a phase transition (the exact identification of which itself is uncertain), it is this situation that in applications of superconductivity requests immediate action. It is this situation that cannot be covered by standard stability models.

But if time cannot be specified exactly and uniquely, how can stability of superconductors be predicted and a quench, initially local, be avoided with safety from spreading over the total conductor cross section? Not only stability predictions become indefinite but also the meaning of critical temperature (and perhaps also of the other critical parameters) of a superconductor.

This situation is not very convincing and is contrary to standard description of the superconductor state: Critical temperature after disturbances, within finite periods of time, is not uniquely defined (either is a non-equilibrium quantity, or it is an equilibrium quantity that cannot be reached within finite time), and analogue conclusions most probably apply to also the superconductor materials parameter $J_{Crit}$ (it is not clear whether this applies to also the critical field, $B_{Crit}$). Estimation of the length of these periods, for the given YBaCuO 123 thin film conductor, can be attempted by inspection of Figure 5b.

As a consequence, it is not possible to uniquely separate local zero loss, flux flow and Ohmic resistances in the conductor cross section and over its lengths within finite periods of time. This means that also distinction of current limiting into flux flow and Ohmic principles, again within finite periods, and description of current percolation, is not clear.

In addition, also gravitating quantum systems can lead to an indefinite space-time metric. Since the Schrödinger equation relies on a time parameter, too, what then plays the role of such time parameter in the absence of a definite metric?

Further, a central question arises from the applied methods and findings reported in this paper remains: Can a generalised "non-transparency" be specified for the four scenarios

- radiation in multi-component heat transfer,
- quantum mechanical entanglement (the EPR experiment),
- localisability of images on time scales,
- loss of individuality of photons if they simultaneously hit
- a beam splitter (the HOM experiment).

in a way that it is relevant for superconductors? Interestingly, non-transparency is the property that makes a thorough and exact analysis of transient superconductor states possible at all.

In the EPR experiment (the instant when particle 2 immediately "knows" that a measurement at particle 1 has been performed), there is no flow of energy running in parallel to a flow of entropy. This situation, and the simulations reported in the present paper apparently have parallels with exchange of information. Accordingly, to which extent is "non-transparency" an information problem?

It presently appears as if the more traditional information gets lost the closer the system superconductor approaches its phase transitions.

Experimental verification of all these predictions could be attempted by measuring a time dependence of the increased electrical conductivity in the so called thermal fluctuations state, of levitation height (during approach to an equilibrium, levitation height should increase) and persistent currents (here immediately after start of the experiment). Observed variations, if any, would be very small.

***************

Final note by the author: All the above is not new physics but presents corollaries and consequences from multi-component heat transfer in superconductor filaments and thin films. It is an attempt to reveal limitations of present limited knowledge and to improve understanding of correlations between superconductor stability and phase transitions.

Comments are welcome.

## 10   References

1   Wilson M N, Superconducting Magnets, in: Scurlock, R. G. (Ed.) Monographs on cryogenics, Oxford University Press, New York, reprinted paperback (1989)

2   Dresner L, Stability of Superconductors, in: Wolf St. (Ed.) Selected topics in superconductivity, Plenum Press, New York (1995)

3   Seeger B. (Ed.), Handbook of Applied Superconductivity, Vol. 1, Institute of Physics Publishing, Bristol and Philadelphia, IOP Publishing Ltd (1998)

4   Reiss H, A microscopic model of superconductor stability, J. Supercond. Nov. Magn. **26** (2013) 593 - 617

5   Reiss H, Inhomogeneous temperature fields, current distribution, stability and heat transfer in superconductor 1G multi-filaments, J. Supercond. Nov. Magn. **29** (2016) 1449 - 1465

6   Reiss H, Radiative Transfer, Non-transparency, Stability against Quench in Superconductors and Their Correlations, J. Supercond. Nov. Magn. **32** (2019) 1529 -1569

7   Reiss H, Stability of a (2G) Coated, Thin Film YBaCuO 123 Superconductor - Intermediate Summary, J. Supercond. Nov. Magn. (2020), https://doi.org/10.1007/s10948-020- 05590-3

8   Reiss H, The Additive Approximation for Heat Transfer and for Stability Calculations in a Multi-filamentary Superconductor - Part A, J. Supercond. Nov. Magn. **32** (2019) 3457 - 3472

9   Reiss H, The Additive Approximation for Heat Transfer and for Stability Calculations in a Multi-filamentary Superconductor - Part B, J. Supercond. Nov. Magn. **33** (2020) 629 – 660

10  Castro-Ruiz E, Giacomini F, Belenchia A, Brukner C, Quantum clocks and the temporal localisability of events in the presence of gravitating quantum systems, Nature Communications| (2020) 11: 2672|https://doi.org/10.1038/s41467-020-16013-1| www.nature.com/naturecommunications

11  Reiss H, Stability considerations using a microscopic stability model applied to a 2G thin film coated superconductor, J. Supercond. Nov. Magn. **31** (2018) 959 - 979

12  Glover III R E, Superconductivity above the transition temperature, in: Gorter C J (Ed.), Progress in low temperature physics, Vol. VI, Chap. 7, North-Holland Publishing Company, Amsterdam, London (1970) 291 - 332

13  Reiss H, Finite element simulation of temperature and current distribution in a superconductor, and a cell model for flux flow

resistivity – Interim results, J. Supercond. Nov. Magn. **29** (2016) 1405 – 1422

14  Siegel R, Howell J R, Thermal radiation heat transfer, Int. Stud. Ed., McGraw-Hill Kogakusha, Ltd., Tokyo (1972)

15  Chandrasekhar S, Radiative Transfer, Dover Publ. Inc., New York (1960)

16  Reiss H, Troitsky, O Yu, Radiative transfer and its impact on thermal diffusivity determined in remote sensing, in: A. Reimer (Ed.), Horizons of World Physics **276** (2012) Chapter 1, pp. 1 – 67

17  Carslaw H S, Jaeger J C Conduction of Heat in Solids, 2$^{nd}$ Ed., Oxford Science Publ., Clarendon Press, Oxford (1959), reprinted (1988), 256 and 356

18  Withaker St, Fundamental Principles of Heat Transfer, Pergamon Press, Inc., New York (1977)

19  Hong C K, Ou Z Y, Mandel L, Measurement of subpicosecond time intervals between two photons by interference, Phys. Rev. Lett. **59** (1987) 2044 -2046

20  Phelan P E, Flik M I, Tien C L, Radiative properties of superconducting Y-Ba-Cu-O thin films, Journal of Heat Transfer, Transact. ASME **113** (1991) 487 - 493

21  Kumar A R, Zhang Z M, Boychev V A, Tanner D B, Vale L R, Rudman D A, Far-Infrared transmittance and reflectance of $YBa_2Cu_3O_{7-\delta}$ films on Si substrates, Journal of Heat Transfer, Transacts. ASME **121** (1999) 844 – 851

22  Viskanta R, Heat transfer by conduction and radiation in absorbing and scattering materials, Transacts. ASME, J. Heat Transfer **2/65** (1965) 143 - 150

23  Reiss H, A short-time physics problem in superconductivity: Stability against quench, Proc. 12th Int. Workshop Subsecond

Thermophysics (IWSSTP2019), Cologne, June 2019, High Temperatures - High Pressures (2020), in print.

**Figures 1 to 9**

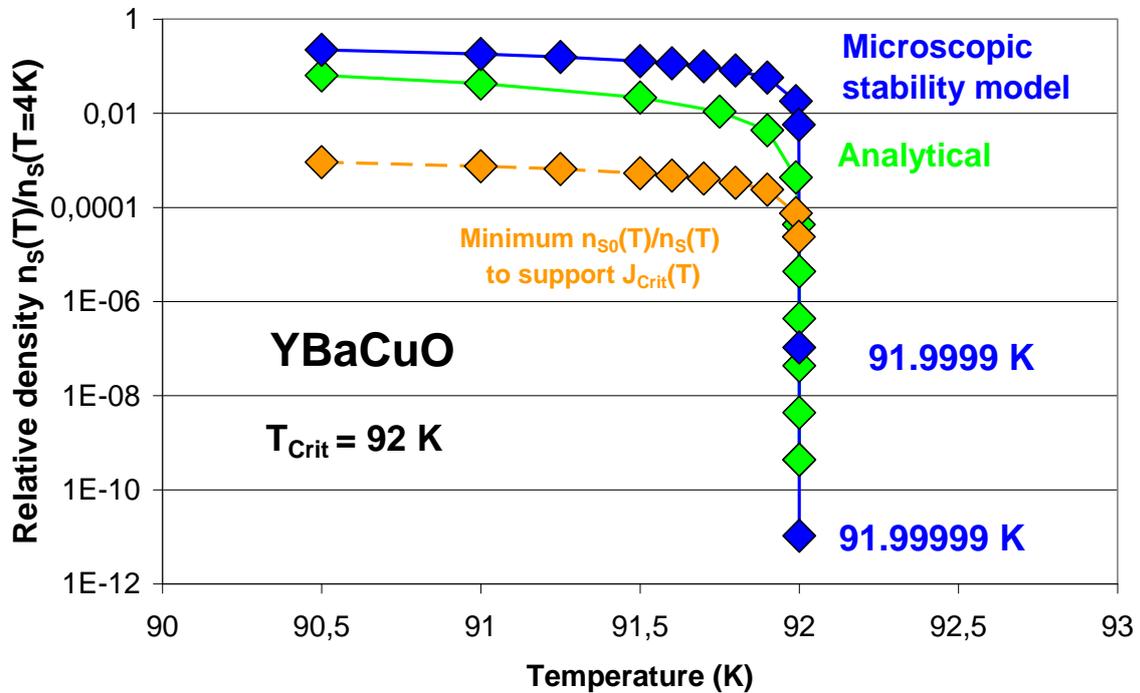

Figure 1a Relative density (the order parameter), $f_S = n_S(T)/n_S(T=4K)$, in dependence of temperature, calculated for the thin film, YBaCuO 123 superconductor (dark-blue diamonds). Dark-yellow diamonds indicate the minimum relative density of electron pairs that would be necessary to generate a critical current density of $3 \cdot 10^{10}$ A/m$^2$ (YBaCuO) at 77 K, in zero magnetic field. The diagram compares predictions of the microscopic stability model with analytical results (light-green) calculated from Eq. (8) in [20]. The Figure is copied with slight modifications from Figure 11b of [6].

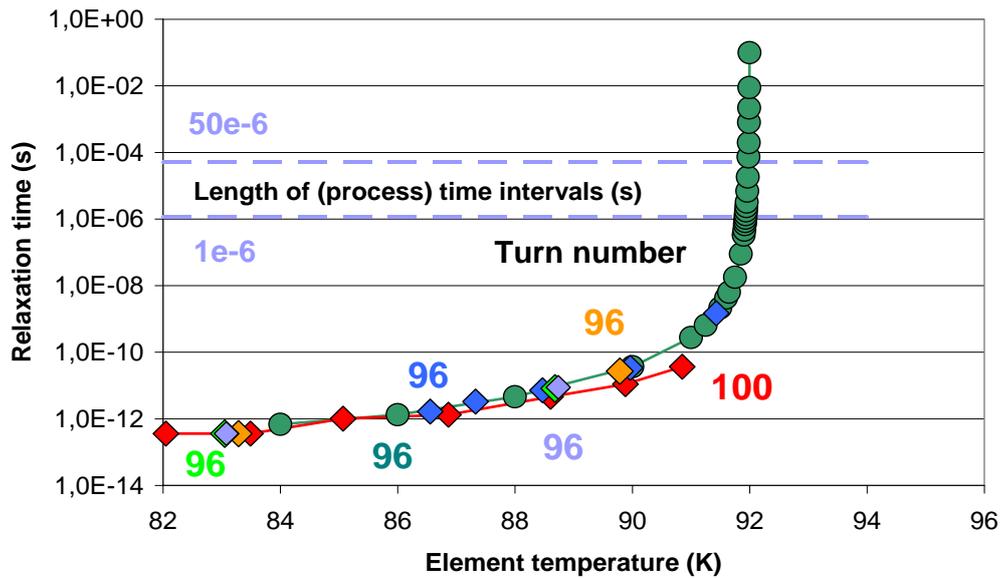

<u>Figure 1b</u> Relaxation time (the time needed to obtain a new dynamic equilibrium after a disturbance, arising at the indicated temperatures in the centroids of turns 96 (light-green, lilac, orange and blue diamonds, respectively) and 100 (red diamonds). The turns belong to a coil prepared from a coated YBaCuO 123 thin film superconductor of standard architecture. Superconductor film thickness is 2 µm. The disturbance originates from transport current density locally exceeding critical current density (the corresponding flux flow losses steadily increase local conductor that may finally become larger than $T_{Crit}$). As soon as Finite Element temperature exceeds 91.925 K, coupling of all electrons in this thin film superconductor to a new dynamic equilibrium can no longer be completed within integration times of 1 or 50 µs. The Figure is copied von Figure 8 of [11].

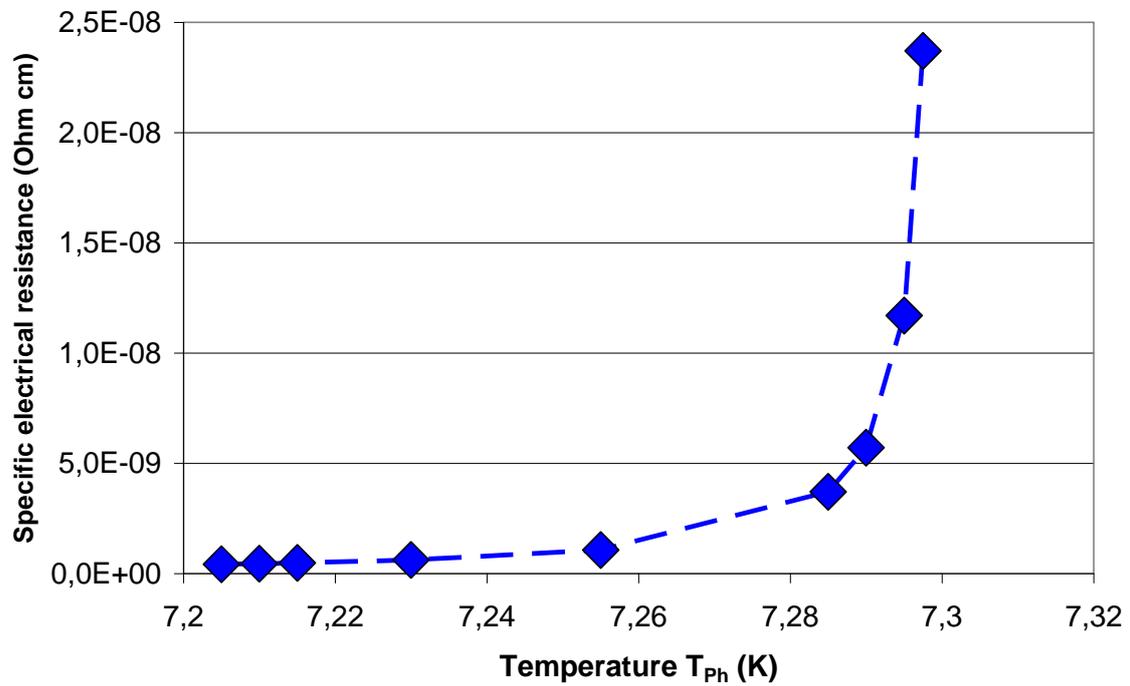

Figure 1c Effective specific electrical resistance, $\rho_{El}$, of Pb in the thermal fluctuations state, $T > T_{Crit}$ (7.2 K). Results are calculated using the Russell cell model. The resistance decays if temperature, $T_{Ph}$, approaches $T_{Crit}$ from values above. At $T = 7.3$ K, the porosity of the normal charge carriers (the part that really competes with the superconducting charge carriers to current transport) is set to $\pi = 1$ in this model (the contribution of the superconducting charge carriers then is zero).

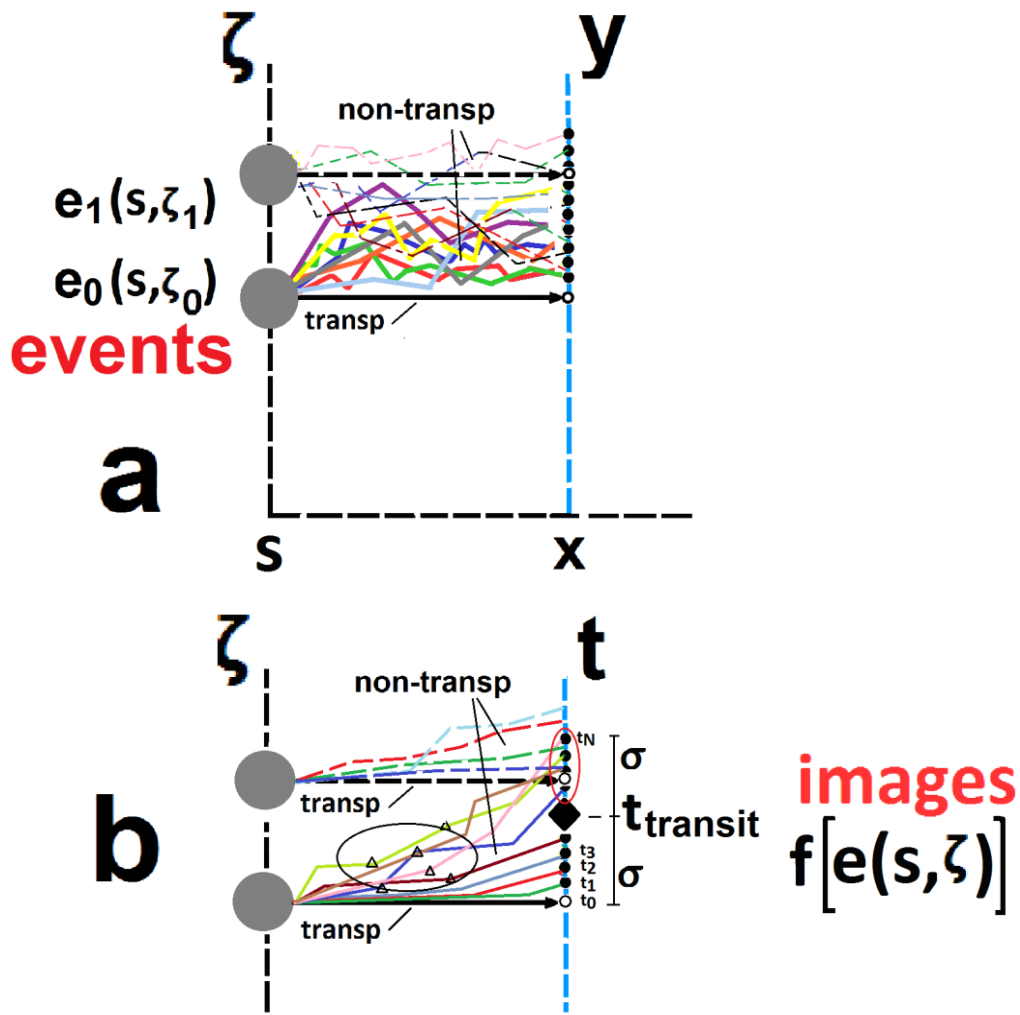

Figure 2a,b The Figure explains mapping functions that under solely radiative transfer in a non-transparent material generate a series of hits and of images, $f[e(\mathbf{s},\zeta_1)],...,f[e(\mathbf{s},\zeta_N)]$, of single events, here $e_1(\mathbf{s},\zeta_1)$ and $e_2(\mathbf{s},\zeta_2)$, like temperature variations. (a) Between the vertical axes s and x, the Figure schematically shows radiative transit paths from the two events (dark-grey circles) that under multiple scattering hit a plane at positions $(x,y_{i,j})$, with $1 \leq i,j \leq N$. Within the same region of space, (b) explains transit times as images of the two events.

In a *non-transparent* sample, the number of images on the time scale, t, generated by each of the events is arbitrarily large (N →∞; small, black solid circles on the time scale, t). The black ellipse in part (b) encloses inner positions (small open triangles) where the derivative of $f[e(\mathbf{s},\zeta)]$ increases or decreases according to statistical variations of the Albedo, $\Omega$, against its mean value; the derivative always exceeds zero. Variations of $\Omega$ are responsible for variations of the scattering part of the extinction coefficient, which means $\Omega$ is also responsible, within the radiative contribution to total heat transfer, for variations of transit time. The red ellipse in (b), again in a non-transparent sample, encloses images like f[e$_1$(**s**,ζ$_1$)] = t$_{1,i}$ and f[e$_2$(**s**,ζ$_2$)] = t$_{2,j}$ of e$_1$(**s**,ζ$_1$) and e$_2$(**s**,ζ$_2$), that become intermixed on the time scale. In parts (a) and (b) of this Figure, intermixing of (a) hits and of (b) images destroys correlation of the order between events and their images. Images generated by conduction heat flow would be overlaid onto the radiative images shown in this Figure.

In case of *transparent* samples, correlations provided by the mapping functions are bijective (thick black, solid or dashed horizontal lines, respectively); for a single event there is one and only one image for each of the events (here indicated by the open circles at t$_0$ for e$_1$ and e$_2$). An infinitely (uncountably) large number of images, like in a dense set, is required for generating a time scale at position **x** or at other coordinates. This works only if the medium is transparent.

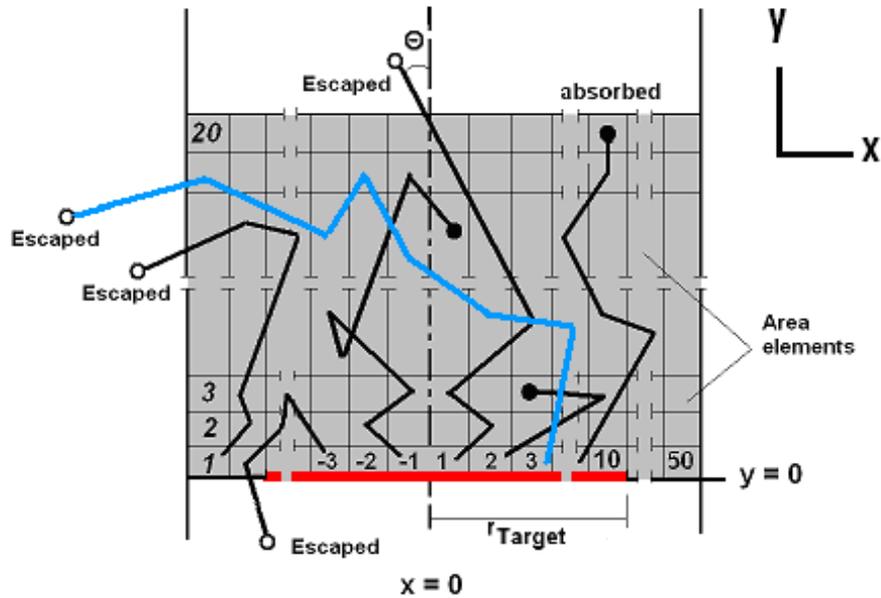

Fig. 3a Coordinate system of a superconductor, thin film. Sample radius and thickness of the film are 5 and 2 µm, respectively. The Figure (schematic, not to scale) shows area elements, k(i,j), radiation bundles (thick solid lines) and the target (thick horizontal red line, target radius 1 µm, small against sample radius). Volume elements are generated by rotating the area elements against the symmetry axis (dashed-dotted line). Scattering angle is denoted by θ. The scheme is used for both Finite Element (FE) and Monte Carlo (MC) calculations. Bundles may escape from the sample (index "Escaped") after a series of absorption/remission or scattering interactions. The blue bundle, as an example originally emerging from a position x > 0 finally escapes from positions x < 0 to neighbouring materials or to coolant; it shows that FE and MC simulations cannot be restricted to only one of the half-sections (we do not have axial symmetry).

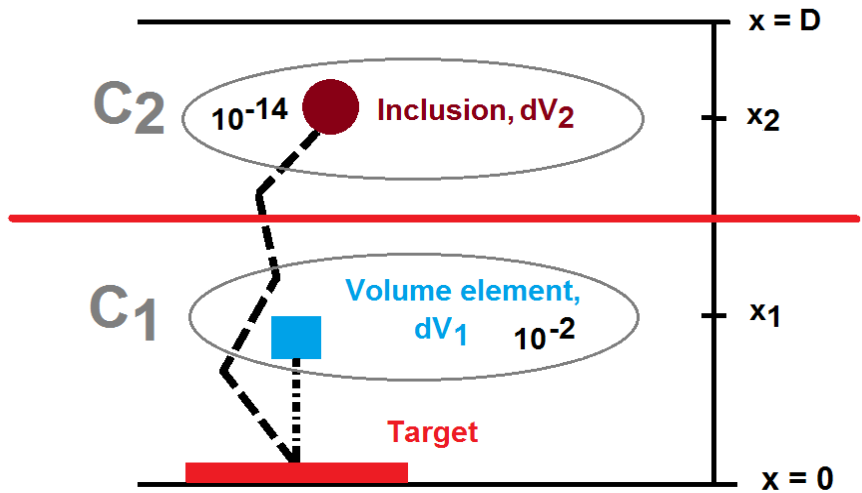

Figure 3b The Figure explains incompleteness of standard Radiative Transfer in non-transparent materials. Regular materials element ($dV_1$, Albedo $\Omega < 1$) and a foreign, strongly absorbing inclusion ($dV_2$, $\Omega \ll 1$), respectively, are indicated by the light-blue rectangle and the dark-brown, solid circle, respectively. Scattered part of a beam (dashed line, emitted from the target) hits the inclusion at an arbitrary depth position, $x_2$, after about $10^{-14}$ s (compare Figure 7a), which causes a local temperature increase within $dV_2$. Absorbed intensity of the same beam (dashed-dotted line), with its absorption coefficient, $A = (1 - \Omega) E$, being much larger than the scattering coefficient, $S = \Omega E$, causes a temperature increase of $dV_1$ at position $x_1$, but at about $10^{-2}$ s after emission. Temperature excursions at both positions originate from the *same* basic event (emission of a beam from the target). In the non-transparent material, both generated temperature excursions, $dT/dt$, when considering the *integral* radiative process, are strictly separated in time (the thick, horizontal red line) if the geometrical distance $x_1 - x_2$ is large against a multiple 15 of the mean free paths of photons (the optical thickness between both positions, $\tau \geq 15$). An observer at $x_1$ does not know what happens at $x_2$ as long as solid conduction, as a *differential* process, from $x_1$ to $x_2$ does not couple the $dT/dt$. As a consequence, there is no common time scale, t, in the "cells" $C_1$ and $C_2$. But solid conduction is a considerably slow process (signals emitted from the target need about $10^{-7}$ s to arrive at $x_2$).

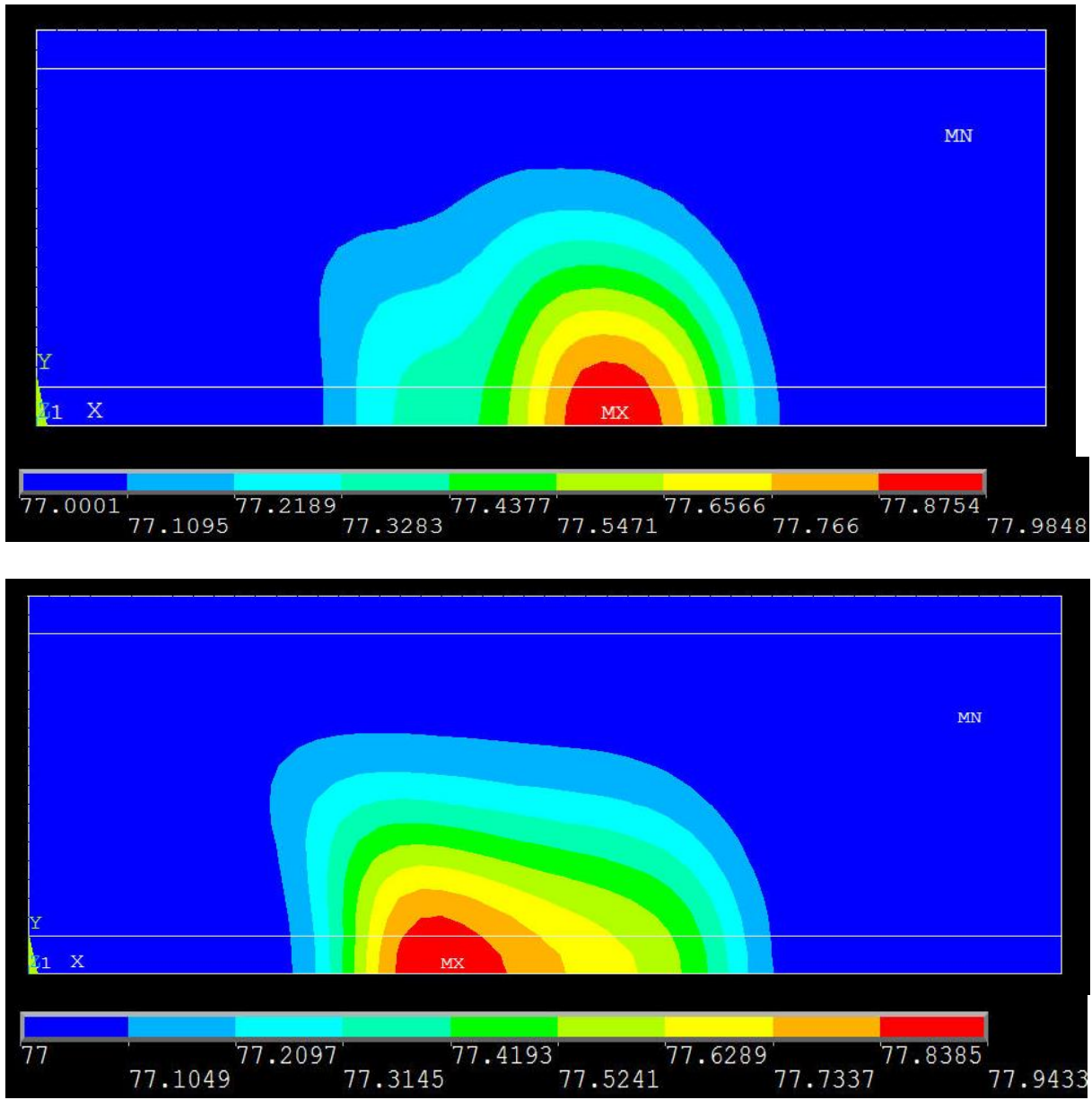

Figure 4a,b Temperature fields within the superconductor sample (the 2 µm YBaCuO 123 thin film). In this Figure and in Figures 4c,d and 5a,b and 6a, in order to enhance the effect exerted by anisotropic thermal conductivity on calculated temperature fields, the crystallographic c-axis has been oriented parallel to the (horizontal) sample x-axis. Results are obtained for the case *"Solid conduction plus radiation"*, which means from Finite Element (FE) with integrated Monte Carlo (MC) simulations, here shown for $t = 5 \cdot 10^{-8}$ s simulation time. The calculations apply a short, weak (rectangular) pulse (a "disturbance") of, in this Figure, in total $Q = 5 \cdot 10^{-12}$ Ws that is incident onto the target area during $0 \leq t \leq 8 \cdot 10^{-9}$ s. A Monte Carlo simulation to generate conductive and radiative heat sources to apply the Carslaw and Jaeger theorem is overlaid onto the target excitation. All heat transfer mechanisms (solid conduction,

radiation) in total are treated as a conduction process; this is specified as the *standard procedure* (FE, MC) as in our previous reports. Like in Figure 5a of [7], sample thickness is divided into 0.1 µm thick boundary layers and the 1.8 µm core (see Subsect. 4.3.2 of [7] for justification). The different temperature distributions in Figure 4a (above) and 4b (below) result from positions, within the target area, of arbitrarily assigned cluster points around which the emission of the Monte Carlo beams are distributed (right or left to sample symmetry axis, respectively). Angular distribution of the residual, directional intensity of the beams, when released at the sample rear side surface, in both cases is isotropic.

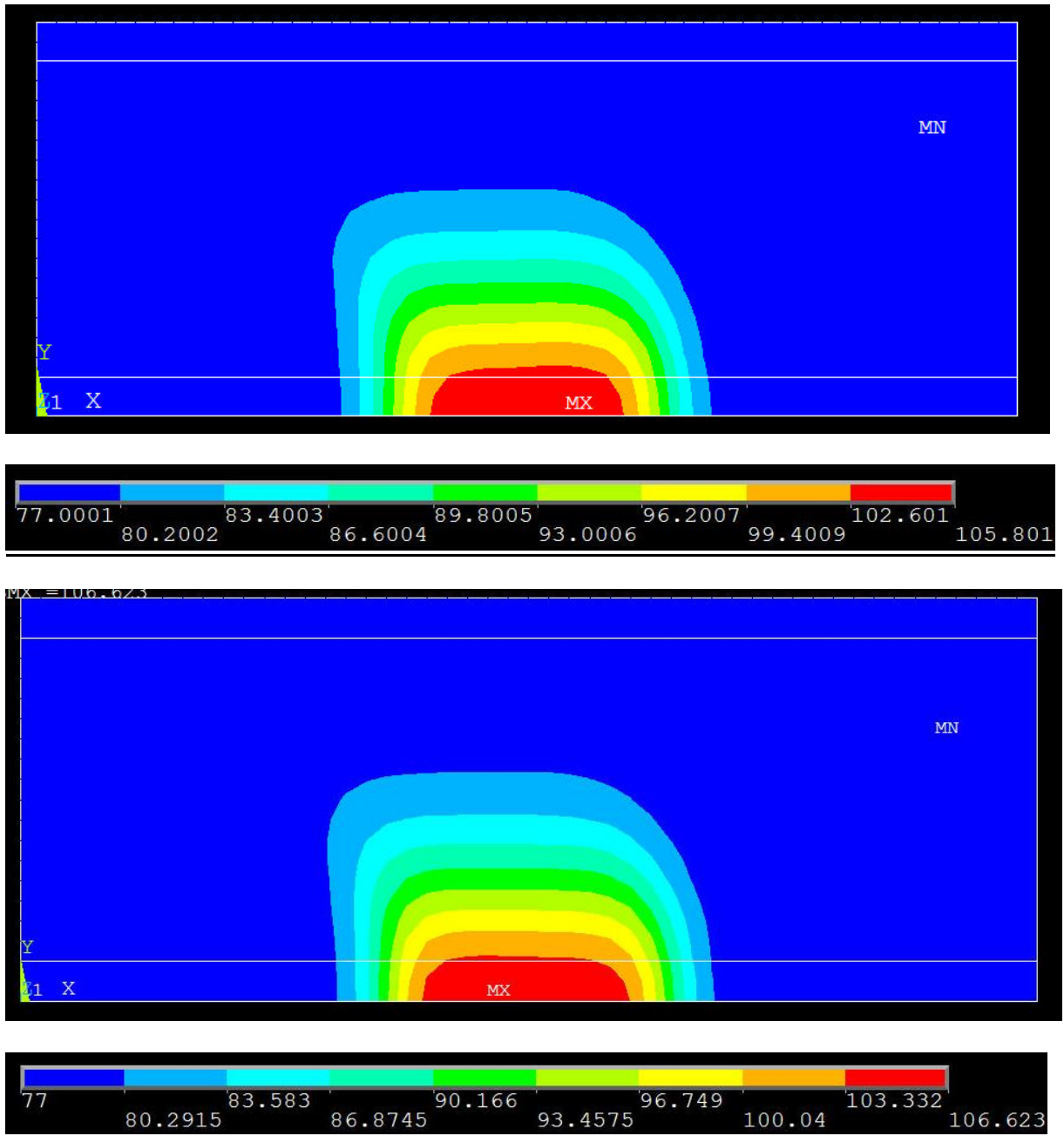

Figure 4c,d Temperature fields within the superconductor sample (the 2 µm YBaCuO 123 thin film). Same calculation as in Figure 4a,b but with the increased load $Q = 2 \cdot 10^{-11}$ Ws. Results are shown as before at $t = 5 \cdot 10^{-8}$ s simulation time. Due to the higher load Q, the difference between both temperature distributions load has become smaller; this results from the increased number of beams that survive the absorption/remission processes. But compare the asymmetric position of the index "MX relative to the sample symmetry axis that denotes temperature maximum.

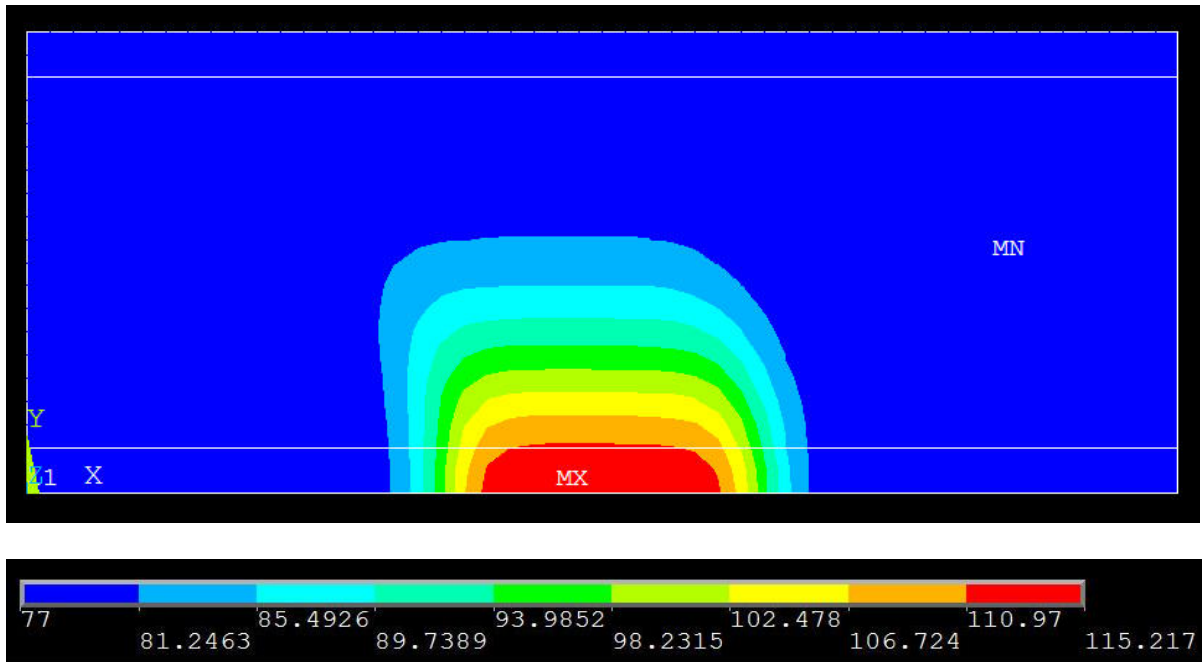

Figure 5a Same calculation as in Figure 4c,d, again with the increased rectangular heat pulse of in total $Q = 2 \cdot 10^{-11}$ Ws, applied to the target area during $0 \leq t \leq 8 \cdot 10^{-9}$ s, but for *solely solid conduction* (obtained from only the FE result). Simulation time as before is $5 \cdot 10^{-8}$ s. The significant temperature increase in comparison to Figure 4c,d results from the condition that part of the heat pulse is not distributed by a Monte Carlo simulation to the total thin film volume but would be applied to only the target surface ($y = 0$).

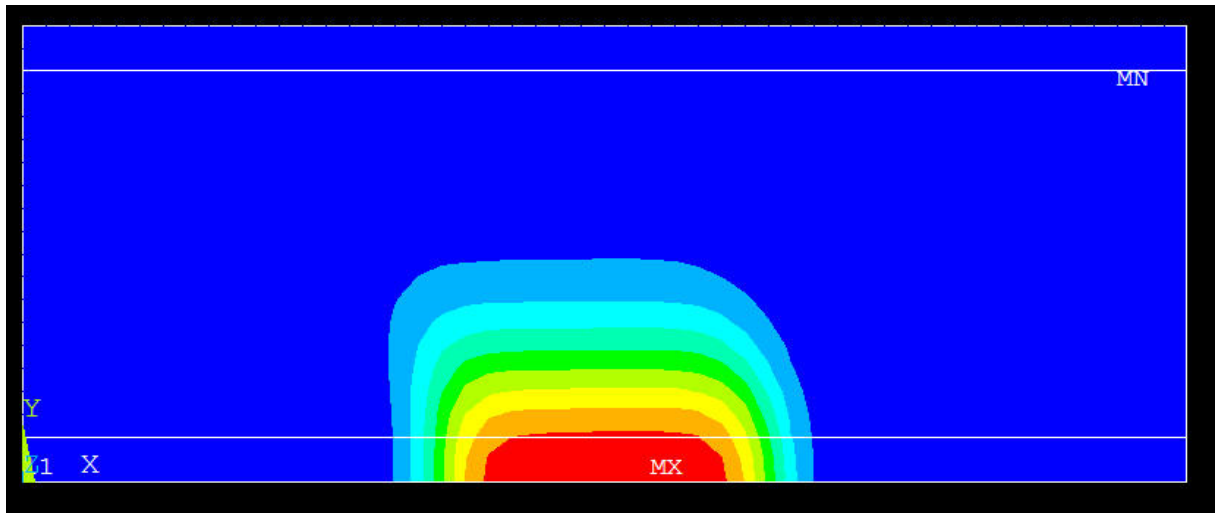
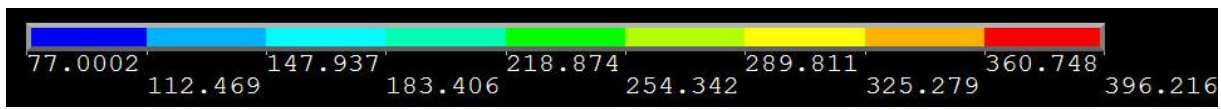

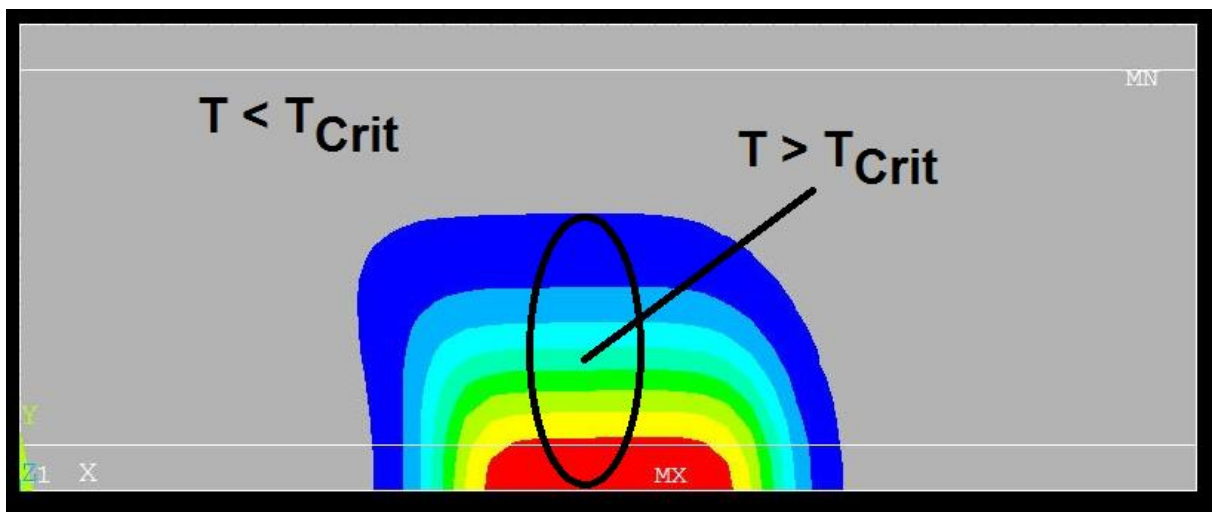
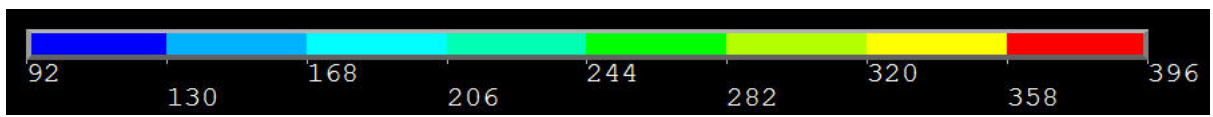

Figure 5b Same calculation as in Figure 5a but for solid conduction plus radiation (using the *standard procedure*, with FE, MC) but for a rectangular heat pulse of in total $Q = 2\ 10^{-9}$ Ws applied to the target during $0 \leq t \leq 8\ 10^{-9}$ s. Simulation time is $5\ 10^{-8}$ s. Results (above): full range of temperatures, and below: for all $T > T_{Crit} = 92$ K. Temperature of a considerable part of the superconductor cross section exceeds $T_{Crit}$. Zero loss current transport would be possible only within the shaded (grey) area.

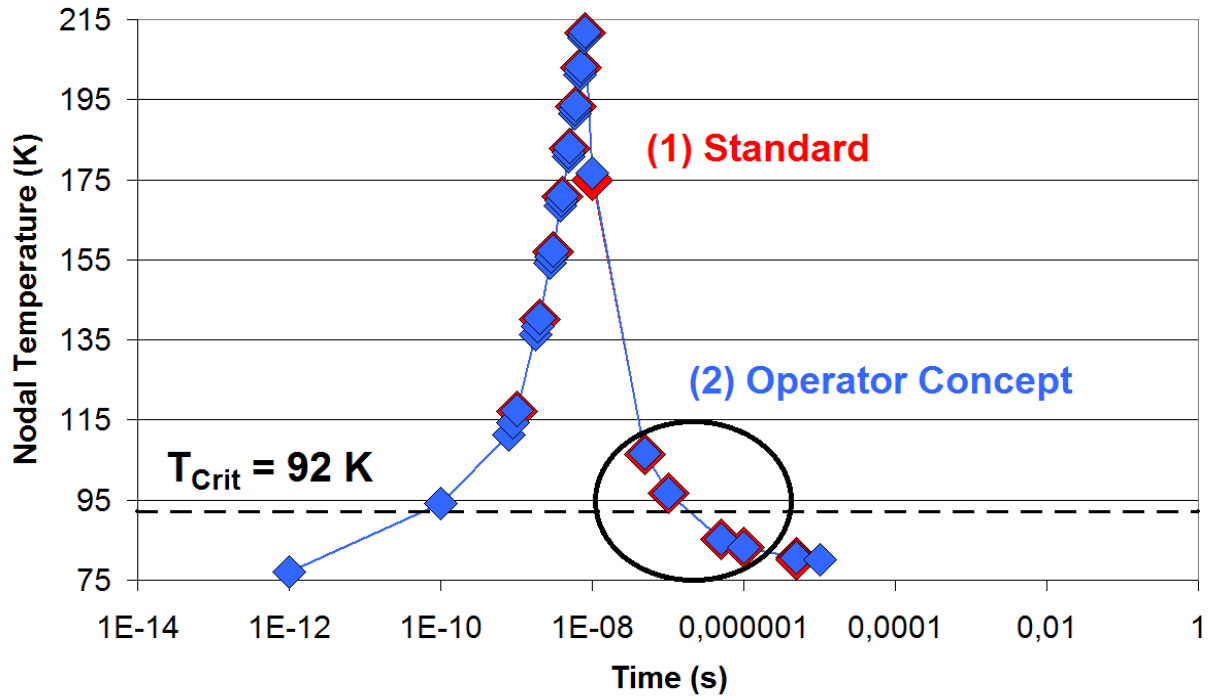

<u>Figure 6a</u> Transient nodal temperature at x = 0, y = 0 (the target centre of the the D = 2 µm YBaCuO 123 thin film). A rectangular heat pulse (the "disturbance" Q) of in total $2 \cdot 10^{-11}$ Ws, is applied to the target area) during $0 \leq t \leq 8 \cdot 10^{-9}$ s. Orientation of the crystallographic c-axis like in the previous Figures is parallel to the sample x-axis. The two curves show (small) differences (high-lighted by the black ellipse) between (1) solid and radiative heat transfer (both in the non-transparent medium) simulated as solely conduction processes (red diamonds, the standard procedure, with no division of the time axis), and (2) the operator concept (light-blue diamonds) using the standard procedure but additionally with intervals time,j and dtime,j ($1 \leq j \leq 4$) as given in the text; the times time,j are applied to divide each load step (nlast) into intervals, with $\Omega = 0$ in $0 \leq t \leq t_1$). In (1) and (2), as in the previous Figures, the original excitation is at the target surface and, using the Monte Carlo simulation, with radiative and conductive heat transfer from sources generated within the sample (this is done to apply the Carslaw and Jaeger theorem, compare [17]). Like in Figures 4a-d and 5a,b, the curves are calculated using the Additive Approximation. Within the load steps, length of the intervals is given by dtime,1 = $10^{-12}$, dtime,2 = $8 \cdot 10^{-10}$, dtime,3 = $9 \cdot 10^{-10}$ and dtime,4 = $10^{-9}$ s. Length of integration time steps within these intervals is $10^{-13}$, $10^{-10}$, $10^{-11}$ and $10^{-11}$ s, respectively.

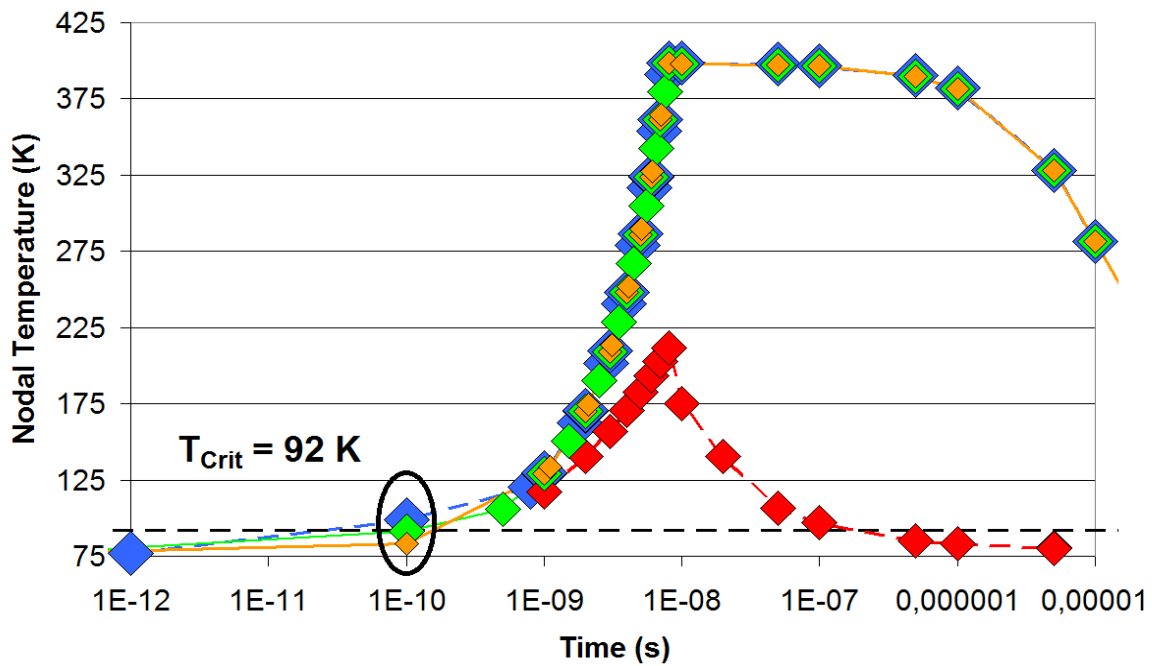

Figure 6b Same calculation as in Figure 6a for the YBaCuO 123 thin film (D = 2 µm) using the operator concept (dark-blue, light-green and dark-yellow diamonds); red diamonds for comparison show the result obtained with the standard procedure, curve (1) in Figure 6a. In this calculation, we have applied in intervals 1 to 3 only radiative conduction while in interval 4, solid conduction is overlaid onto radiation. The small radiative conductivity is responsible for strongly increased conductor temperature (it acts like a thermal mirror). Increasing weight is assigned to temporal contributions of solid conduction by decreasing value of dtime,3 (from 0.9 to 0.5 and 0.1, given by the dark-blue, light-green and dark-yellow diamonds, respectively). Reduction of dtime,3, because of the much larger solid conductivity, accordingly results in a continuous decrease of nodal temperature, compare the diamonds in the black ellipse. The uncertainty of solid temperature seen around $T_{Crit}$, here in the early stages of temperature excursion after the disturbance, Q, may question superconductor stability. At later times, data points almost coincide but then $J_{Crit}$ is zero, in any case, and have no impacts on stability against quench.

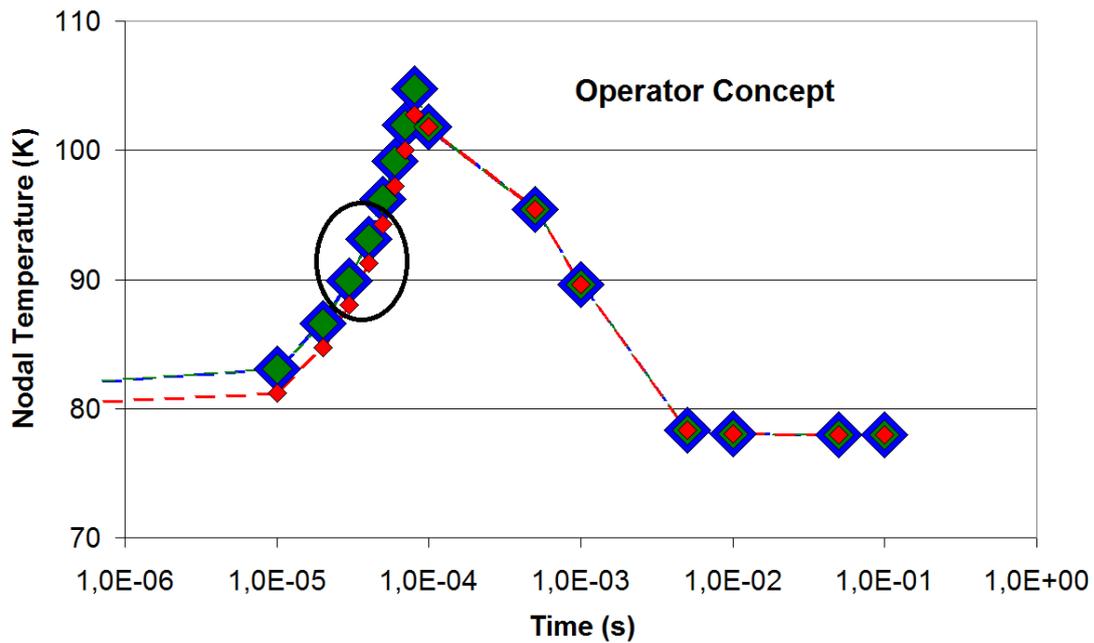

Figure 6c Transient nodal temperature at x = 0, y = 0 (solid, dark-blue diamonds). Same calculation as in Figure 6a using the operator concept but with increased length of the load steps (each $10^{-5}$ s) and with orientation of the c-axis (anti-) parallel to the sample y-axis. Total incident energy onto the target here is Q = 2 $10^{-10}$ Ws. The differences between the dark-green and red diamonds reveal the effect of strong (by one magnitude) parameter variations (anisotropy ratio, χ, of the solid thermal conductivity in ab- and c-axis direction) made during the sequence of time intervals (but orientation of the c-axis remains fixed). In particular within the black circle, temperature variations are seen to become increasingly important (against the results in Figure 6a,b) for stability analysis. The Figure again is calculated using the Additive Approximation. Values of dtime,j (1 ≤ j ≤ 3) in this calculation are identical, dtime,j = $10^{-13}$ s, which results in strong contribution by solid conduction, and accordingly reduced conductor temperature in relation to Figure 6b. Length of integration time steps within these extended intervals is $10^{-14}$, $10^{-14}$, $10^{-14}$ and $10^{-6}$ s.

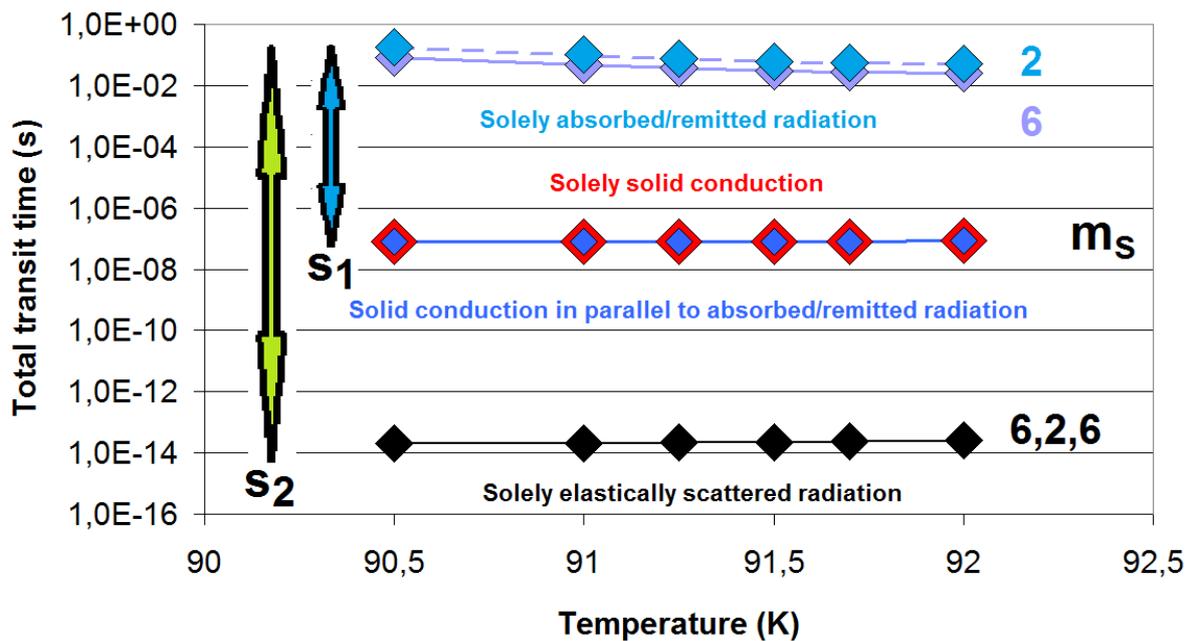

Figure 7a Transit time of signals (events) proceeding by different heat transfer mechanisms through the 2 µm YBaCuO 123 thin film. The sequence in time of heat transfer mechanisms after an event is specified as (1) solely elastic scattering, (2), (3) and (4) solid conduction in parallel to absorbed/remitted radiation (still under scattering, but scattering does not contribute to temperature excursion) The possible case of in-elastically scattered radiation is not considered. Cases (2) to (4) are calculated from the diffusion approximation $L = C (a_{Th} t)^{0.5}$ using anisotropy factors, $m_S$, indicated in the Figure, and the Albedo $\Omega$ (taken from Figure 14b of [6]). Time spans, $s_1$, denotes time lag between two, fundamentally different, separate heat transfer mechanisms (solely solid conduction, radiation), $s_2$ is the maximum difference between two completely different, radiative transport processes.

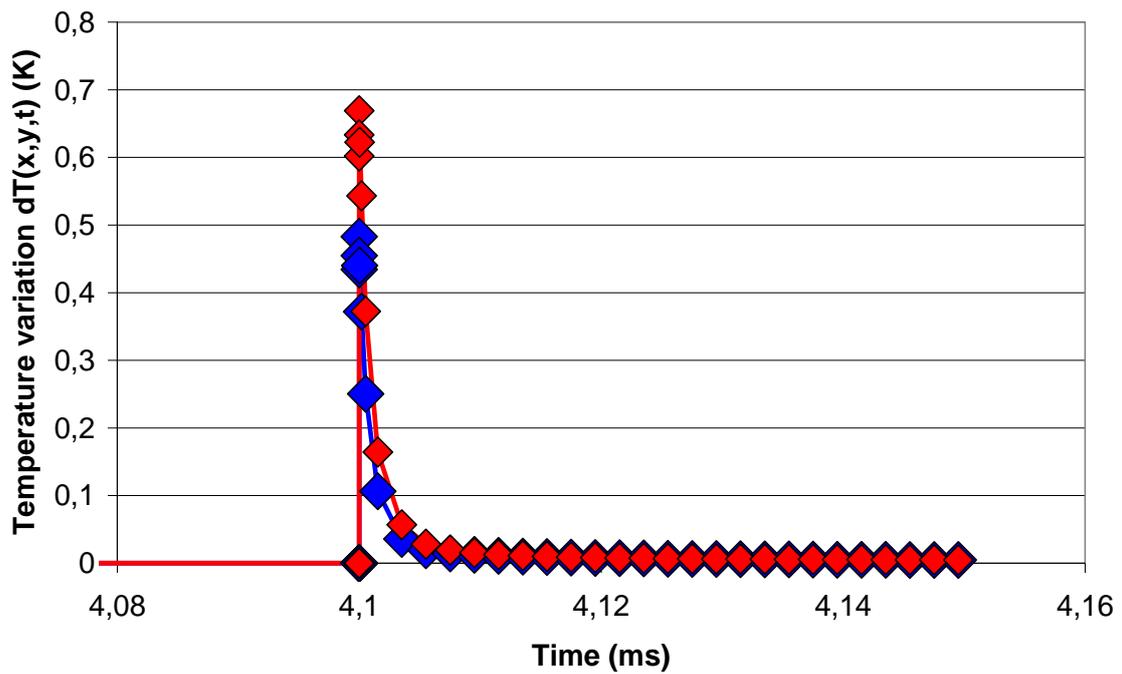

Figure 7b Variation dT(x,y,t) of temperature of the "second generation" (2G) coated, YBaCuO 123 thin film superconductor during its transit time under solid conduction parallel to radiation (solely absorption/remission), in the YBaCuO 123 thin film (the coil [11], turn 96). Results are given for positions close to the axis of symmetry (red diamonds), and at the outermost left thin film position of the thin film (blue diamonds).

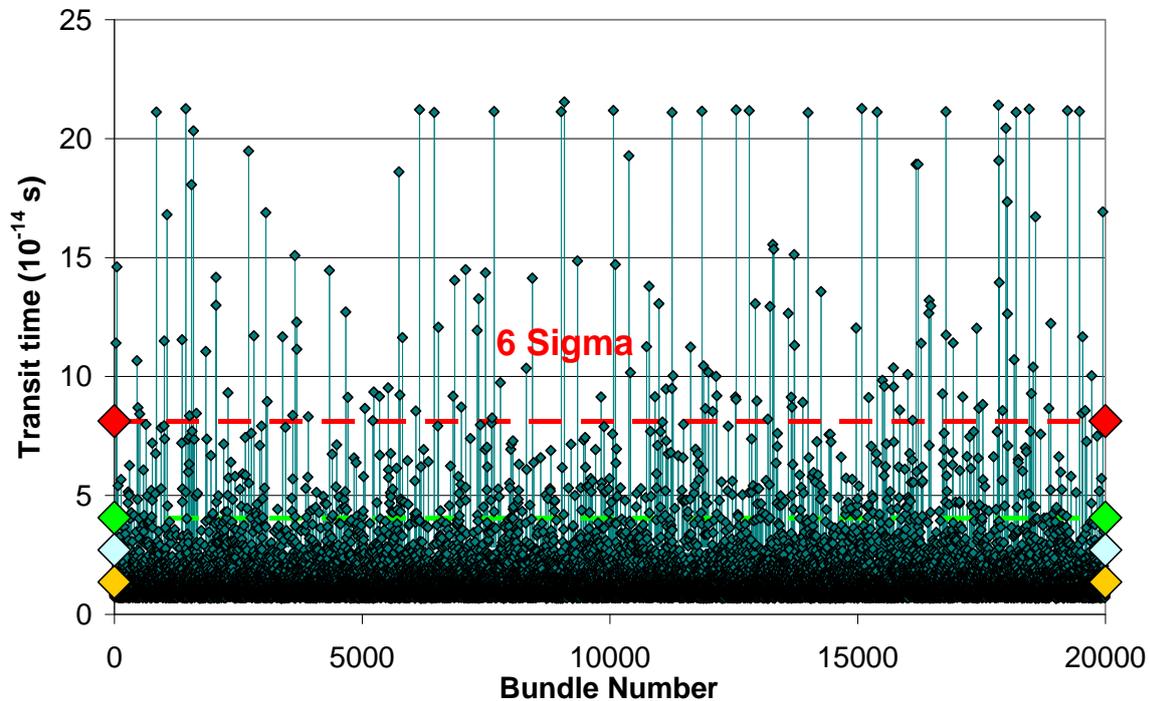

Figure 8a Monte Carlo simulation of transit time, $t_{Trans}$, of individual bundles, again in the 2 µm "second generation" (2G) coated, YBaCuO 123 thin film superconductor, here under pure elastic scattering. Results are shown for the first $2 \cdot 10^4$ (of in total $M = 5 \cdot 10^4$) bundles. Separately for the thin, 0.1 µm boundary layers and the 1.8 µm core of the thin film (Figure 5a of [7]), the calculations apply extinction coefficients for dependent and independent scattering, $m_S$-factors (for forward scattering) and Albedo, all from [6]). The yellow, light-blue, light-green and red diamonds and the dashed lines indicate 2- or 3- or 4- or 6-sigma, safety intervals, respectively.

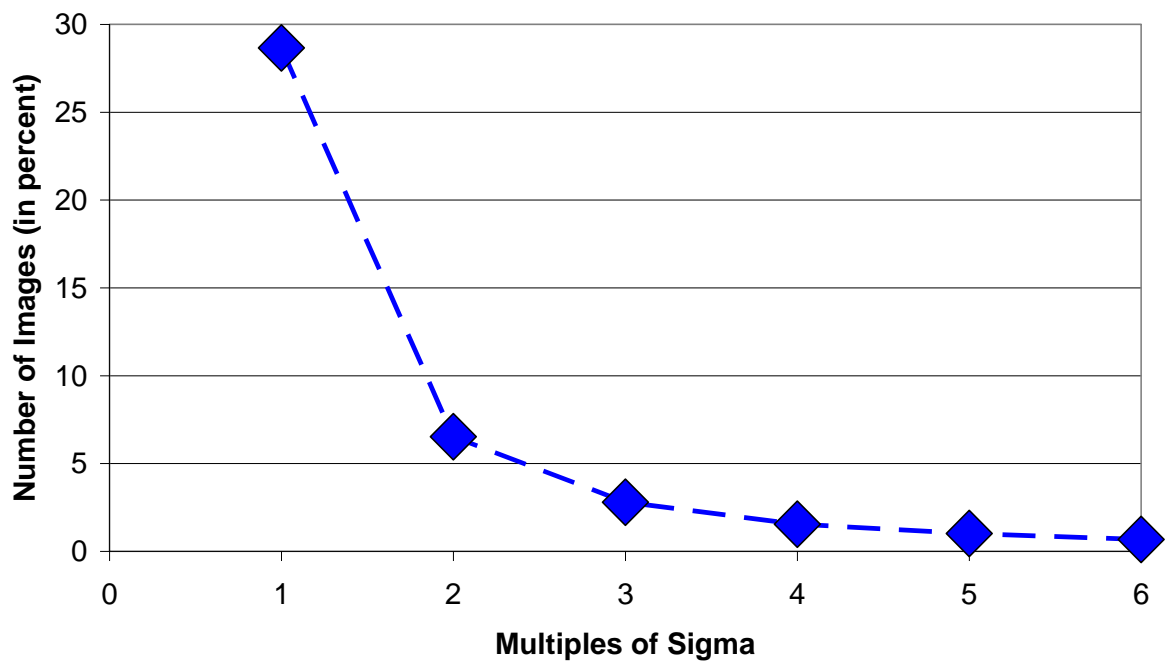

<u>Figure 8b</u> Number of images (bundles, given in per cent of the total M = 5 10$^4$) that do *not* fall into multiples between 1σ to 6σ of the data shown in Figure 10a,b of [7].

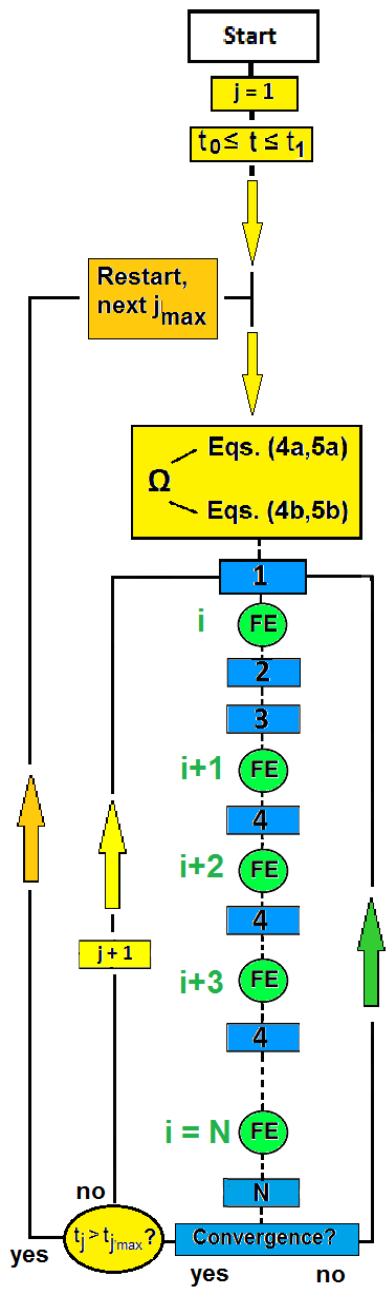

Figure 9 Flow chart (the "master" scheme) showing two iteration levels (i, j) and one time loop ($t_j$) of the numerical simulation: Light-green circles and indices, i: Sub-steps, the proper Finite Element (FE) calculations; Light-yellow indices, j: Load steps involving FE and, within the blue rectangles, critical current, magnetic field and resistance (flux flow, Ohmic) calculations; Dark-yellow indices, t: Time loop, lines of the matrix **M** (equation numbers shown in the rectangles refer to [7]). The blue rectangles with sub-step numbers i = 1, 2, 3,...N are defined as **1:** First FE step, j, with data input of start values of temperature distribution, specific resistances, critical parameters of J, B and of initial (uniform)

transport current distribution or of single, isolated radiation heat pulses, respectively; **2:** Results obtained after the first FE step (i), if converged, for the same parameters in the *same* load-step, j; calculation of $T_{Crit}$, $B_{Crit}$, $J_{Crit}$; **3:** Calculation of resistance network and of transport current distribution (if applicable), all to be used as data input into the next FE calculation (sub-step i + 1), within the *same* load step, j; **4:** Results like in **2**; Sub-steps **5, 6,...N:** Results like in **3** or **4;** convergence yes or no ? If "no", return to **1** (iteration i = 1, in the same load step, j). If "yes" go to next load step j + 1, continue with **1**. The number N of FE calculations (green circles) might strongly increase computation time. Length of simulation time, $t \leq t_{max}$, within each of the individual intervals, with $t_{max}$ indicating the maximum time of a corresponding particular interval, is selected according to the different transit times, source functions, different radiation propagation mechanisms, different ratios of solid conduction and radiation and to different wavelengths. More description of the FE steps can be found in Sect. 2.3 of [11] and for the possible correlation between non-convergence and quench in Figure 1 of [9]. By the time-loop, t, the Figure is an extension of Figure 12 of [9]. For an example, see Figure 14 of [8].

# Appendices 1 to 4

## Appendix 1: Iterative Numerical Procedure

In order to relieve the reader from repeatedly consulting details in our previous papers, the following explains the principle how the simulation of the stability problem have been performed.

An iterative "master" scheme (Figure 9) has been applied that incorporates Finite Element (FE) simulation load-steps to calculate transient temperature, $T(x,y,t)$, from Fourier's differential equation. When convergence has been achieved, the scheme calculates electrical and magnetic superconductor states, critical values of temperature, magnetic induction and of current density, and the distribution of transport current density in the conductor cross section. All these are calculated as local, transient values. The Meissner state is checked in each element of the FE mesh and in each load-step.

Details of the Finite Element part of the scheme (selection of FE elements, meshing, solvers, time steps, convergence criteria) are described in our previous papers.

The simulations assume the superconductor (e. g. a coil or a current limiter) is integrated in a standard, low or medium voltage system that includes Ohmic and induction resistances.

## Appendix 2: Non-uniformity of Temperature Distributions

Under transient disturbances, superconductor temperature neither is uniform in the total superconductor/matrix cross section nor is it uniform within filaments of multi-filamentary superconductors. As an example, temperature distributions are calculated for the 1G BSCCO 2223/Ag

Long Island multi-filamentary superconductor under a sudden fault current.

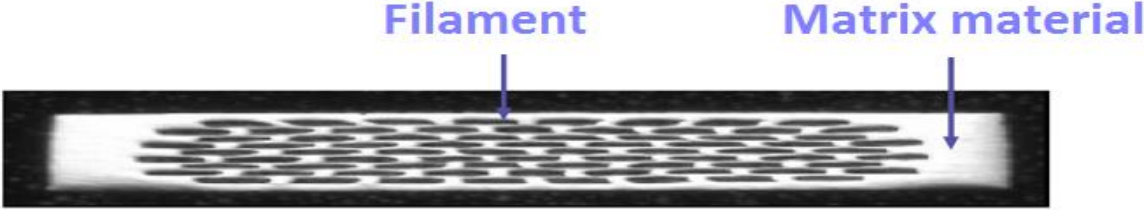

Figure 10 a Full cross section of the 1G BSCCO 2223/Ag Long Island multi-filamentary superconductor. The Figure is taken from Marzahn (2007).

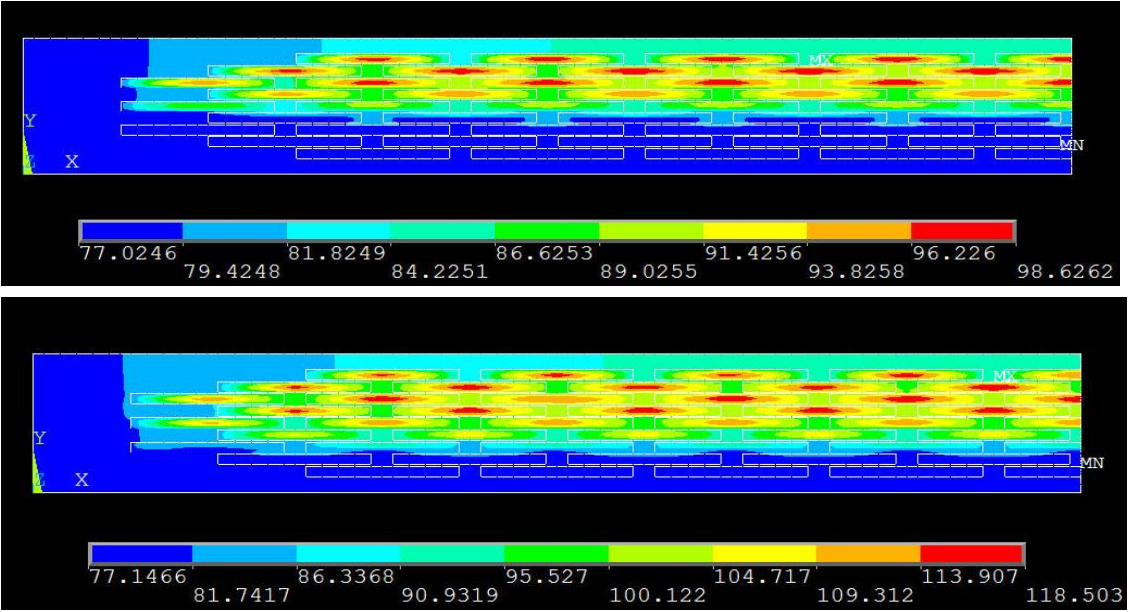

Figure 10b,c Temperature distribution (nodal temperatures) in the cross section of the multi-filamentary conductor (Figure 10a) shown at t = 1.8 ms (above, with all temperatures below critical temperature, $T_{Crit}$ = 108 K at zero magnetic field), and at t = 2.1 ms after start of the disturbance (bottom), a sudden increase, within 2.5 ms, of AC transport current to a multiple of 20 times its nominal value. Because of symmetry, only the left half of total conductor cross section needs to be shown in the two diagrams. Figure 10a-c is copied from previous work of the author.

# Appendix 3: Radiation parameters and examples for coupled conductive/radiative heat transfer and for the Stability Function

| Radiation temperature (K) | 90,5 | 91 | 91,25 | 91,5 | 91,6 | 91,7 | 91,8 | 91,9 |
|---|---|---|---|---|---|---|---|---|
| Wave number (1/cm) | 312,3 | 314,0 | 314,9 | 315,8 | 316,1 | 316,4 | 316,8 | 317,1 |
| Wave length (µm) | 32,02 | 31,84 | 31,76 | 31,67 | 31,64 | 31,60 | 31,57 | 31,53 |
| Real part n of m | 7,21 | 9,81 | 11,34 | 13,16 | 14,01 | 14,97 | 16,11 | 17,58 |
| Imaginary part k | 60,57 | 60,02 | 59,77 | 59,55 | 59,48 | 59,42 | 59,38 | 59,37 |
| Extinction coefficient ($10^7$ 1/m) | 1,409 | 1,446 | 1,469 | 1,497 | 1,510 | 1,525 | 1,543 | 1,566 |
| Albedo | 0,91 | 0,88 | 0,87 | 0,85 | 0,85 | 0,84 | 0,83 | 0,82 |

Table 1 Radiation parameters used for calculation of the radiative conductivity, $\lambda_{Rad}$, applied in the Finite Element simulations of temperature fields in the YBaCuO 123, thin film superconductor. Real and imaginary parts of the refractive index, $m = n - ik$, are obtained from application of the Drude formulas to experimental data for the complex dielectric constant, $\varepsilon$, reported by Kumar et al. [21]. Scattering cross sections and extinction coefficients, given here for independent scattering, and Albedo of single scattering, are obtained from application of rigorous scattering (Mie) theory using spectral values (taken at maximum of the Black Body radiation curve) of the complex refractive index, m. Compare [6] for details of the calculations.

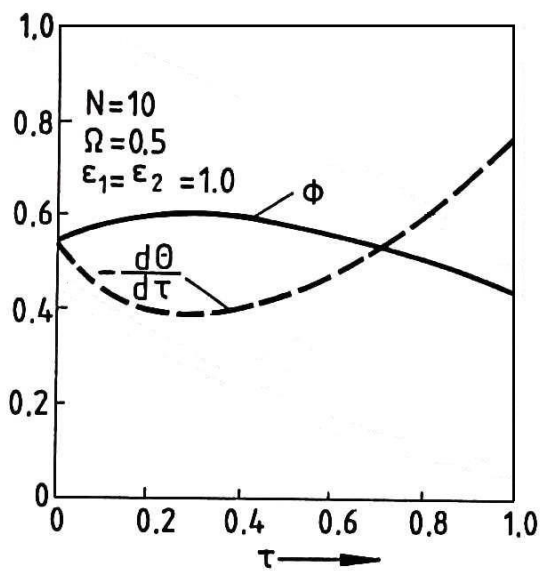

Figure 11 Viskanta's results [22] obtained for coupled conductive/radiative heat transfer in a homogeneous, conductive, absorbing/remitting and isotropically scattering sample (the material is assumed to be grey). The curves are given as function of local optical thickness, τ, and for different $N = 4\, n^2\, \sigma\, T^4/(\lambda_{Cond}\, E\, T_1)$, using n the refractive index, σ the Stefan-Boltzmann constant, $\lambda_{Cond}$ the thermal conductivity, E the extinction coefficient, $T_1$ the temperature at the "hot wall" (τ = 0), Ω the Albedo, $\varepsilon_{1,2}$ the thermal emimssivities of side walls 1 and 2 (all constant), and the dimensionless temperatures, $\Theta_{1,2} = T_{1,2}/T_1$. Content of the original Figure has been reduced to only the curves for -dΘ/dτ and Φ. The optical parameters are integral (not spectral) values.

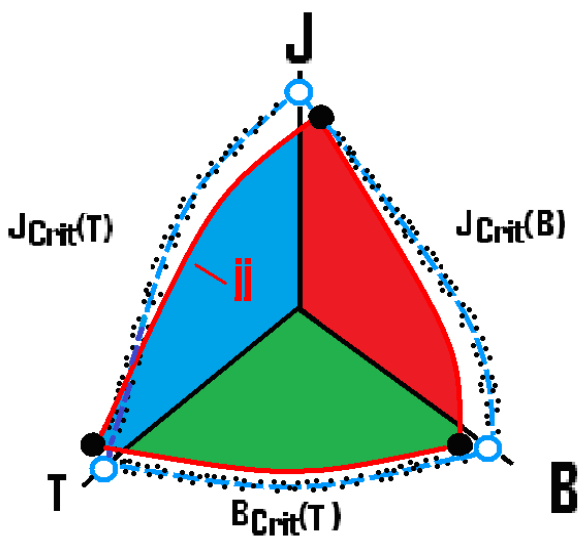

Figure 12 Existence diagram of type II superconductivity (schematic, not to scale; the lower critical magnetic field is not shown). The dashed blue line and the open blue circles in this Figure denote the *conventional* region of existence of superconductivity. Random variations $\Delta T_{Crit}$, $\Delta B_{Crit,2}$ and $\Delta J_{Crit}$ against conventional values, here of YBaCuO 123, indicated by small black dots, shall account for shortcuts in conductor manufacturing and handling; this applies to the existence diagrams of all elements in the Finite Element scheme. The random variations (taken in each of the Finite Elements) are within ± 1 K, ± 5 Tesla and ± 1 percent, respectively. The Figure is copied from [23].

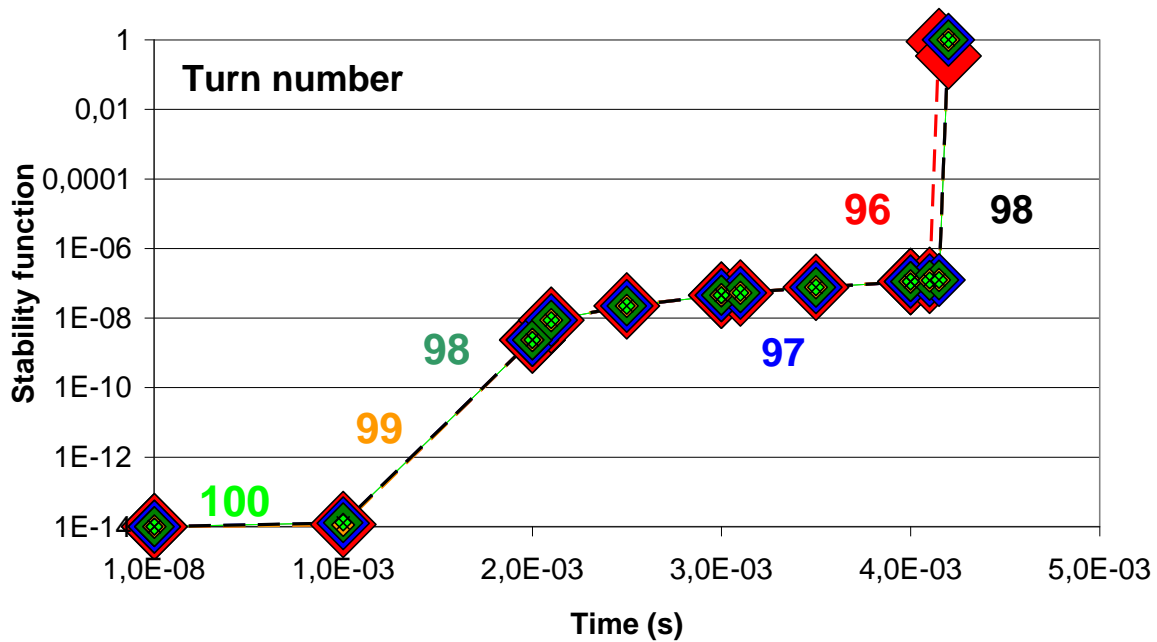

Figure 13 Stability function, Φ(t), obtained for solid conduction plus radiation (and solid/liquid heat transfer at the solid/coolant contacts) in the thin film, YBaCuO 123 superconductor. The Figure is copied from [7]. Results are shown for turns 96 to 100 of the coated conductor winded to a coil, as described in this reference. The calculations assume a sudden increase of transport current above its nominal value beginning at t = 3 ms.; flux flow resistances then are responsible for thermal losses that locally increase conductor temperature. All curves apply the same ("standard") uncertainties $\Delta T_{Crit}$, $\Delta B_{Crit,2}$ and $\Delta J_{Crit}$ of the electrical/magnetic critical parameters $T_{Crit}$, $B_{Crit,2}$ and $J_{Crit}$ (92 K, 240 Tesla and $3\ 10^{10}$ A/m$^2$ at T = 77 K, respectively) and of the anisotropy factor (r = 10, again at T = 77 K) of the thermal diffusivity. Coloured curves are obtained with no random fluctuations of the solid thermal conductivity, $\lambda_{Cond}$ black crosses instead apply to a ± 5 percent random variation of $\lambda_{Cond}$ at random positions within turn 98.

Note the steep increase of the stability function at about 4 ms after start of the simulations. It coincides with the quench.

**Appendix 4: Time scales**

Physical time scales do not exist a priori, they need events for their definition. Time scales cannot exist without events, like space cannot exist without bodies. The following applies to non-relativistic situations.

For a concept to describe properties of time scales in situations close to phase transitions, and to apply these properties in numerical simulations, we assume:

Time scales, i. e. elements incorporated in time scales, can be defined as isomorph to the set **R** of real numbers. Here, we are interested in physical time (its counterpart, psychological time, is not considered). The set **R** cannot be replaced by the set of rational numbers.

The set **R** contains an uncountably large number of elements, which means the distribution of elements within **R** is dense (the property "dense" is explained in the text). This is the pure mathematical content. Physically, the temporal "distance" between two events or between two images cannot be infinitesimally small; there is at least a lower limit given by resolution limits set to practical experiments. The numerical procedure selects only those events or images that can be assigned to the physics of a process under consideration (like a temperature variation, decay of an electron pair, or a local quench). But potentially the whole set of elements of **R** would be available for the simulations. The question whether this applies to only positive elements is left open.

Correlation between an arbitrary physical situation (a system located at a position, **s**, within which an event takes place) and the set **R** is realised by mapping functions that project events, e(**s**,ζ), in their natural order, ζ, onto a time scale. The projections of the events are their images, f[e(**s**, ζ)] = e(**s**,t); these are correlated to elements of **R** where they within **R** are ordered if mapping is bijective; which replaces ζ by t.

Correlations provided by the mapping functions and bijective mapping exist per se, if the system is transparent. Mapping does not require any action exerted by individuals or by mechanisms, it is a purely mathematical construct.

In Figure 14 the decisive steps 1 and 3 correlate events with images on the time scale, t, that in turn are bijectively correlated with (mapped onto) the set **R**. The set **R** in this Figure is expanded to **R**$^3$ to explain that the correlations between events and images can be considered at arbitrary positions **x** and **x'** in the superconductor cross section or volume.

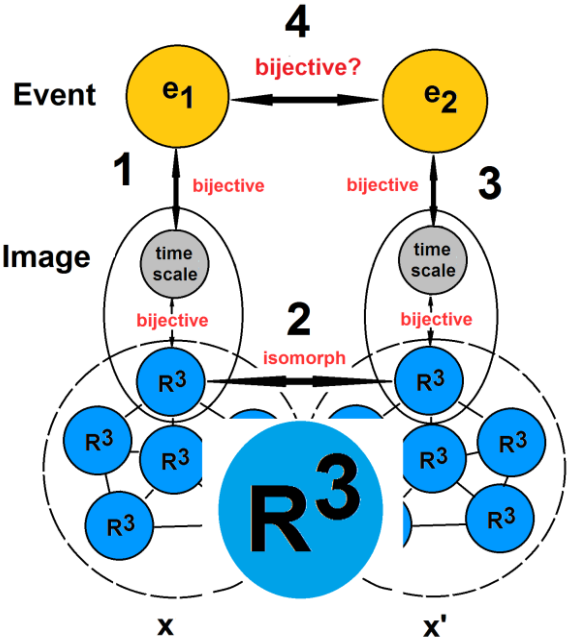

Figure 14 A cycle (schematic) to explain coupling between events, $e_1$ and $e_2$, their images, time scales and the set of real numbers, **R**. Time scales are coupled, by bijective mapping, to an infinitely large number of sets $\mathbf{R}^3$, wherein all sets are isomorph to each other (the dashed circles thus encloses elements the number of which is infinitely large and is yet identical to $\mathbf{R}^3$). The cycle consisting of steps (1) to (4) can be completed with bijective mapping functions and isomorphisms only in case the medium is transparent (which means, if also relation 4 is bijective).

The complete cycle is the identity (if correlation 4 is bijective). Once bijective mapping between time scales and the set $\mathbf{R}^3$ is accepted, the set $\mathbf{R}^3$ takes the role as a "vehicle". In non-transparent media, relation (4) in Figure 14 is not uniquely defined. If distances of $\mathbf{x}'$ from $\mathbf{x}$ are smaller than optical thickness of the medium, exceptions might be possible.

These conclusions apply the following lemma:
<u>Lemma</u>: A single time scale to exist implies existence of an arbitrarily large number of other time scales.

A proof can be attempted by considering the construction of the set, $\mathbf{R}^3$, of real numbers from four axioms: Trichotomy, transistivity, compatibility with both addition and multiplication. No any other automorphism apart from identity is allowed in $\mathbf{R}^3$. Time scales, by definition, shall be isomorph to $\mathbf{R}^3$. Therefore, an arbitrarily number of isomorph sets $\mathbf{R}^3$ and, referring to the initial assumption (see above), an arbitrarily large number of time scales, can be generated, the latter only if the medium is transparent.

Temporal localisability relies on transparency, temporal non-localisability of events accordingly results from non-transparency. A question remains: Is time itself transparent?